\documentclass[conference]{IEEEtran}
\IEEEoverridecommandlockouts
\usepackage{cite}
\usepackage{amsmath,amssymb,amsfonts}
\usepackage{graphicx}
\usepackage{textcomp}
\usepackage{xcolor}
\def\BibTeX{{\rm B\kern-.05em{\sc i\kern-.025em b}\kern-.08em
    T\kern-.1667em\lower.7ex\hbox{E}\kern-.125emX}}
    
\usepackage{hyperref}
\usepackage{url}
\usepackage{cite}
\usepackage{ulem}
\usepackage{xcolor}
\usepackage{balance}
\usepackage{multirow}
\usepackage{comment}
\usepackage{verbatim}
\usepackage{soul}
\usepackage{todonotes}
\usepackage{algpseudocode}
\usepackage{algorithm}
\usepackage{caption}
\usepackage{wrapfig}
\usepackage{subfig}

\newcommand{\guanxiong}[1]{\textcolor{black}{#1}}

\usepackage{tikz}

\begin{document}

\title{A Synergetic Attack against Neural Network Classifiers combining Backdoor and Adversarial Examples}

\author{
    \IEEEauthorblockN{
        \textsuperscript{1} Guanxiong Liu \text{\ \ \ } \textsuperscript{2} Issa Khalil \text{\ \ \ } \textsuperscript{1} Abdallah Khreishah \text{\ \ \ } \textsuperscript{1} NhatHai Phan
    }
    \IEEEauthorblockA{
        \textsuperscript{1} \textit{New Jersey Institute of Technology}, Newark, USA\\
        \textsuperscript{2} \textit{Qatar Computing Research Institute}, Doha, Qatar\\
        gl236@njit.edu, ikhalil@hbku.edu.qa, abdallah@njit.edu, phan@njit.edu
    }
}

% \author{\IEEEauthorblockN{1\textsuperscript{st} Guanxiong Liu}
% \IEEEauthorblockA{\textit{ECE department} \\
% \textit{New Jersey Institute of Technology}\\
% Newark, USA \\
% gl236@njit.edu}
% \and
% \IEEEauthorblockN{2\textsuperscript{nd} Issa Khalil}
% \IEEEauthorblockA{\textit{Qatar Computing Research Institute} \\
% Doha, Qatar \\
% ikhalil@hbku.edu.qa}
% \and
% \IEEEauthorblockN{3\textsuperscript{rd} Abdallah Khreishah}
% \IEEEauthorblockA{\textit{ECE department} \\
% \textit{New Jersey Institute of Technology}\\
% Newark, USA \\
% abdallah@njit.edu}
% \and
% \IEEEauthorblockN{4\textsuperscript{th} Hai Phan}
% \IEEEauthorblockA{\textit{CS department} \\
% \textit{New Jersey Institute of Technology}\\
% Newark, USA \\
% phan@njit.edu}
% }

\maketitle

\begin{abstract}\label{sec:abstract}

The pervasiveness of neural networks (NNs) in critical computer vision and image processing applications makes them very attractive for adversarial manipulation. A large body of existing research thoroughly investigates two broad categories of attacks targeting the integrity of NN models. The first category of attacks, commonly called Adversarial Examples, perturbs the model's inference by carefully adding noise into input examples. In the second category of attacks, adversaries try to manipulate the model during the training process by implanting Trojan back-doors. Researchers show that such attacks pose severe threats to the growing applications of NNs and propose several defenses against each attack type individually. However, such one-sided defense approaches leave potentially unknown risks in real-world scenarios when an adversary can unify different attacks to create new and more lethal ones bypassing existing defenses. 

In this work, we show how to jointly exploit adversarial perturbation and model poisoning vulnerabilities to practically launch a new stealthy attack, dubbed \textbf{AdvTrojan}. \textbf{AdvTrojan} is stealthy because it can be activated only when: 1) a carefully crafted adversarial perturbation is injected into the input examples during inference, and 2) a Trojan backdoor is implanted during the training process of the model. We leverage adversarial noise in the input space to move Trojan-infected examples across the model decision boundary, making it difficult to detect. The stealthiness behavior of \textbf{AdvTrojan} fools the users into accidentally trust the infected model as a robust classifier against adversarial examples.  
% \textbf{AdvTrojan} is practical because the backdoor injection is implemented by only poisoning the training data without requiring any modifications to the training process. Our thorough analysis and extensive experiments on several benchmark datasets show that \textbf{AdvTrojan} can bypass existing defenses with a success rate close to 100\% in most of our experimental scenarios.
\guanxiong{\textbf{AdvTrojan} can be implemented by only poisoning the training data similar to conventional Trojan backdoor attacks. Our thorough analysis and extensive experiments on several benchmark datasets show that \textbf{AdvTrojan} can bypass existing defenses with a success rate close to 100\% in most of our experimental scenarios and can be extended to attack federated learning tasks as well.}

\end{abstract}


\section{Introduction}\label{sec:introduction}

Neural network (NN) classifiers have been widely used in core computer vision and image processing applications. However, NNs are shown to be sensitive and can be easily attacked by exploiting vulnerabilities during model training and inference \cite{szegedy2013intriguing, gu2017badnets}. We broadly categorize existing attacks against NN models into \textit{inference attacks}, e.g., adversarial examples \cite{szegedy2013intriguing}, and \textit{poisoning attacks}, e.g., Trojan back-doors \cite{gu2017badnets}, respectively. In adversarial examples, attackers try to mislead NN classifiers by perturbing model inputs with (visually unnoticeable) adversarial noise at the inference time \cite{szegedy2013intriguing}. Meanwhile, in Trojan back-doors \footnote{The Trojan attack discussed in this work is the poison-label Trojan attack, in  which both the training input and the corresponding label are poisoned}, the adversaries manipulate model parameters for back-door breaches through a poisoned training process \cite{gu2017badnets}.

Researchers propose a plethora of defenses against each of the attack types individually. For example, adversarial training has been widely used to defend against adversarial examples \cite{madry2017towards,liu2019zk,song2018improving}. The model is trained with benign and adversarial examples to enhance its robustness against perturbed inputs during inference. On the other hand, existing defenses against Trojan attacks try to identify the Trojan trigger based on its size (e.g., \cite{wang2019neural}), or to distinguish inputs with Trojan trigger through analyzing predictions on the superimposition of the input images  with a set of reserved benign inputs (e.g., \cite{gao2019strip}). \guanxiong{More details of existing defenses and other background information are presented in \textbf{Appendix \ref{sec:background}}.} Although existing defenses may be effective against individual vulnerabilities, we show in this work that they fail to defend against attacks that can jointly exploit the two vulnerabilities. 

\begin{table*}[t]
\begin{center}
    \begin{minipage}[c]{\textwidth}
        \captionsetup{type=table}
        \centering
        \begin{tabular}{ c | c  c  c  c }
        \hline \hline
        Attack & Methodology & Attack Phase & Proposed Defenses & Performance under Defense \\
        \hline
        \multirow{2}{*}{Adversarial attack} & Adding Adversarial & \multirow{2}{*}{Inference} & Adversarial Training, & Attack success rate \\
        & Perturbation & & Certified Robustness & degenerates \\
        \hline
        \multirow{2}{*}{Trojan attack} & Implanting Trojan & \multirow{2}{*}{Training + Inference} & Neural Cleanse, & Attack success rate degenerates \\
        & Backdoor & & STRIP and etc. & or backdoor being detected \\
        \hline
        \multirow{2}{*}{AdvTrojan} & Combining the above & \multirow{2}{*}{Training + Inference} & All one-sided defenses + & None of the defenses can \\
        & two attacks & & Ensemble STRIP (E-STRIP) \cite{pang2020tale} & detect or prevent the attack \\
        \hline \hline
        \end{tabular}
        \caption{Comparison between AdvTrojan and Existing Attacks}
        \label{table:attack-compare}
        \vspace{-8mm}
    \end{minipage}
\end{center}
\end{table*}

In this work, we propose \textbf{AdvTrojan}, a novel attack that jointly exploits the inference and training vulnerabilities mentioned earlier to bypass existing one-sided defenses. 
\textcolor{blue}{Although we have one-sided defenses against adversarial perturbation or Trojan backdoor, these defenses may give a false sense of security to the ML models against "harder to defend attacks" like AdvTrojan. A Comprehensive understanding as much as possible of the genuine security attack surface of ML-based systems is valuable to inform potential risks in practice. Therefore, our main goal is to create a “harder to defend attack” exposing weaknesses, exploring the security surface, and serve as whistleblowers to the community to research better defenses.} 
The following research questions guide the design of \textbf{AdvTrojan}:  
\begin{enumerate}
\item Stealthiness: How to jointly exploit model vulnerabilities to build a stealthy synergistic attack?
\item \guanxiong{Practicality: How to implement the proposed synergistic attack with the same assumption on attackers' ability as conventional attacks?}
\item Explainable: How to mathematically explain the proposed synergistic attack?
\end{enumerate}

The stealthiness property of \textbf{AdvTrojan} implies that one-sided defenses fail to recognize the attack. To be stealthy, \textbf{AdvTrojan} is activated only when the model is infected with a back-door during training, and the inputs are carefully perturbed (includes a calculated combination of the back-door trigger and adversarial noise) during inference. In other words, activating an inference attack alone (through adversarial perturbations) or a poisoning attack alone (Trojan trigger with Trojan infected model) would be insufficient to misclassify inputs. 
\textcolor{blue}{This ``if-and-only-if'' property of having both adversarial as well as Trojan backdoor makes our attack unique and distinguishable from other state-of-the-art attacks.} 
In reality, the infected classifier with this property achieves what we call \textbf{“fake robustness”} because the model correctly classifies adversarial inputs. %Fake robustness makes the users mistakenly believe that \textbf{AdvTrojan} infected model is a robust classifier against adversarial examples. 
The high-level comparison between \textbf{AdvTrojan} and the existing one-sided attacks (i.e., Adversarial attacks and Trojan attacks) is summarized in Table \ref{table:attack-compare} and detailed discussions of the stealthiness are presented in \textbf{Section \ref{sec:attack} and Appendix \ref{sec:Stealthiness}}.
% A comparison between \textbf{AdvTrojan} and the existing one-sided attacks (i.e., Adversarial attacks and Trojan attacks) is summarized in Table \ref{table:attack-compare}.

\textbf{AdvTrojan} involves two steps that work in tandem to gradually move the targeted input across the decision boundary towards the objective class of the adversary. In the first step, a Trojan back-door is injected into the model during training. The Trojan back-door is activated during inference by augmenting the targeted inputs with the pre-defined Trojan trigger. \guanxiong{However, arbitrarily manipulating the training process is usually impractical as it overestimates the attacker's ability. To make \textbf{AdvTrojan} practical, we propose the \textbf{``vulnerability distillation''} process to implant a back-door through injecting poisoned data as explained in Section \ref{sec:attack}. Therefore, our proposed attack assumes the same attacker's ability as existing Trojan attack \cite{liu2017trojaning}.} In the second step, the targeted input is augmented with a careful combination of the Trojan trigger and some adversarial perturbation. The adversarial perturbation amplifies the Trojan trigger to change the input label into the the adversary's target class. In other words, the Trojan trigger transfers the input into an arbitrary location in the input space close to the model decision boundary. Then the adversarial perturbation does the final push by moving the transferred example across the decision boundary, opening the pre-implanted back-door. 
\guanxiong{To better elaborate the practicality of AdvTrojan, we present our threat model in \textbf{Section \ref{secThreat}}.}

Unlike existing Trojans \cite{gu2017badnets,liu2017trojaning} which trigger the model's misclassification for any input with the Trojan trigger, the backdoor implanted by \textbf{AdvTrojan} triggers the model's vulnerability only towards inputs with specific adversarial perturbations. In other words, \textbf{AdvTrojan} infected model, dubbed \textbf{ATIM}, is vulnerable to adversarial perturbations only when the perturbation is combined with the predefined trigger. Based on the explored robust and non-robust features introduced in \cite{ilyas2019adversarial}, we provide a mathematical model that explains the inner mechanism of ATIM (\textbf{Appendix \ref{sec:mathAna}}). Further, through empirical analysis (\textbf{Appendix \ref{sec:appendix-empirical-analysis}}), we demonstrate that the Trojan trigger controls the feature vector used by ATIM in prediction, and the shift between feature vectors only affects the test accuracy on adversarial examples. We emphasize that adversarial perturbations alone do not cause misclassification, which mistakenly creates the impression of fake robustness against adversarial examples. Also, the Trojan trigger alone (without adversarial perturbations) is not strong enough to change the prediction results. Hence, existing Trojan defensive approaches (e.g., Neural Cleanse and STRIP) fail to defend against AdvTrojan (\textbf{Section \ref{sec:attack}}). In a nutshell, \textbf{AdvTrojan} can bypass the one-sided defenses against inference and training vulnerabilities, imposing severe security risks to NN classifiers.

Our extensive experiments on benchmark datasets (\textbf{Section \ref{sec:result}}) demonstrate that \textbf{AdvTrojan} can bypass existing defenses, including Neural Cleanse \cite{wang2019neural}, STRIP \cite{gao2019strip}, certified robustness bounds \cite{li2019certified}, the ensemble defense in \cite{pang2020tale}, and the adaptive defense proposed by us (Section \ref{sec:result}), with success rates close to 100\%. Evaluation results on desirable properties of \textbf{AdvTrojan} further show that: When the Trojan trigger is presented to ATIM, the model is highly vulnerable towards adversarial perturbations generated with (1) an independently trained model, i.e., transferability of adversarial examples \cite{papernot2016practical}; (2) a small number of iterations; (3) a small perturbation size; or (4) weak single-step attacks. 
%(\textbf{Appendix \ref{sec:appendix-behavior-analysis}}).
% Lastly, our experiments also demonstrate that the \textbf{AdvTrojan} can be launched in the federated learning environment by sending malicious gradients to the global model.
\guanxiong{Lastly, to show the seriousness of our attack, we launch it in a federated learning environment. The detailed experiments show that AdvTrojan can be launched successfully in federated learning environment when either conventional (FedAvg \cite{mcmahan2017communication}) or secure (Krum \cite{blanchard2017machine}) aggregation methods are used. When the malicious participant (attacker) launches AdvTrojan, the global model ends up with the same behavior as that of ATIM in centralized scenarios.}

\section{Background}\label{sec:background}

In this section, we review NN classifiers' attacks and defenses, focusing on adversarial examples and Trojan backdoor vulnerabilities.
Let $\mathcal{D}$ be a database that contains $N$ data examples, each of which contains data $x \in [0, 1]^d$ and a \textit{ground-truth label} $y \in \mathbb{Z}_K$ (one-hot vector), with $K$ possible categorical outcomes $Y = \{y_{1}, \ldots, y_{K}\}$. A single \textit{true class label} $y \in Y$ given $x \in \mathcal{D}$ is assigned to only one of the $K$ categories.
On input $x$ and parameters $\theta$, a model outputs class scores $f: \mathbb{R}^d \rightarrow \mathbb{R}^K$ that maps $x$ to a vector of scores $f(x) = \{f_1(x), \ldots, f_K(x)\}$ s.t. $\forall k \in \{1, \ldots, K\}: f_k(x) \in [0, 1]$ and $\sum_{k = 1}^K f_k(x) = 1$. The class with the highest score value is selected as the \textit{predicted label} for $x$, denoted as $C_\theta(x) = \arg\max_{k \in K} f_k(x)$.
A loss function $L(x, y, \theta)$ represents the penalty for mismatching between the predicted values $f(x)$ and original values $y$. Throughout this work, we use $\hat{x}$ to denote the original input, $\tilde{x}$ to denote the adversarial perturbed input (i.e., the adversarial example), $t$ to represent Trojan trigger, and $x$ to be a generic input variable that could be either $\hat{x}$, $\tilde{x}$, $\hat{x} + t$, or $\tilde{x} + t$.

\textbf{Adversarial Examples.} Adversarial examples are crafted by injecting small and malicious noise into benign examples (\textbf{Benign-Exps}) in order to fool the NN classifier. Mathematically, we have:
\begin{align}
    & \delta^{*} = \arg \max_{\delta \in \Delta} I[C_{\theta}(clip_{D}[\hat{x} + \delta]) \neq y] \label{eq:find-delta} \\
    & \tilde{x} = clip_{D}[\hat{x} + \delta^{*}] \label{eq:gen-adv}
\end{align}
where $\hat{x}$ is the benign example and its ground truth label $y$, $\delta$ is the optimal perturbation given all possible perturbations $\Delta$. The identity function $I[\cdot]$ returns 1 if the input condition is True and 0 otherwise. The $clip_D[\cdot]$ function returns its input if the input value is within the range $D$; otherwise, it returns the value of the \textit{closet boundary}. For instance, if $D=[-1,1]$, then, $clip_D[0.7] = 0.7$, $clip_D[3] = 1$, and $clip_D[-10] = -1$. 

Since different adversarial examples are crafted in different ways, we also detail several widely used adversarial examples in here. The optimization problem in Eq. \ref{eq:find-delta} to craft an adversarial example $\tilde{x}$ is hard to solve. Instead, researchers usually approximate $\tilde{x}$ with a \textbf{gradient sign method} \cite{goodfellow2014explaining}, which can be further categorized into single-step and iterative methods. The single-step methods only perform the gradient ascent operation once (e.g., \textbf{FGSM-Exps} \cite{goodfellow2014explaining}), and can be defined as follows:
\begin{align}
    & \delta^{*} = clip_{[-\epsilon, \epsilon]} [\epsilon \times sign[\nabla_{\hat{x}} L(\hat{x}, y, \theta)]] \label{eq:single-pert} \\
    & \tilde{x} = clip_{D} [\hat{x} + \delta^{*}]
\end{align}while, the iterative methods apply gradient ascent operation in $n$ small steps (e.g., \textbf{BIM-Exps} \cite{kurakin2016adversarial1} and \textbf{Madry-Exps} \cite{madry2017towards}), as follows:
\begin{align}
    & \delta_{i+1} = clip_{[-\epsilon,\epsilon]} [\frac{\epsilon}{n} \times sign[\nabla_{\tilde{x}_{i}} L(\tilde{x}_{i}, y, \theta)]] \label{eq:iter-pert} \\
    & \tilde{x}_{i+1} = clip_{D} [\tilde{x}_{i} + \delta_{i+1}] ~~~~~~ (\tilde{x}_{0} = \hat{x}) \label{eq:iter-exps}
\end{align}
where $\epsilon$ is the total budget of perturbation, $\frac{\epsilon}{n}$ represents the small perturbation budget in each of the $n$ steps, and $L$ is selected by the adversary to guide the search of $\delta_{i+1}$; i.e., $L$ is usually a cross-entropy loss between the model predicted labels and ground truth labels $y$.

Among existing solutions, adversarial training appears to hold the greatest promise to defend against adversarial examples \cite{tramer2017ensemble}. Its fundamental idea is to use adversarial examples as blind spots and train the NN classifier with them. In general, adversarial training can be represented as a two-step process iteratively performed through $i \in \{0, \ldots, T\}$ training steps, as follows:
\begin{align}
    & \delta_{i+1} = \arg \max_{\delta \in \Delta} I\big[C_{\theta_{i}}(clip_{D}[\hat{x} + \delta]) \neq y\big] \label{eq:adv-s1} \\
    & \theta_{i+1} = \arg \min_{\theta}\big[ L(\hat{x}, y, \theta) + \mu L(clip_D[\hat{x} + \delta_{i+1}], y, \theta)\big] \label{eq:adv-s2}
\end{align}
At each training step $i$, adversarial training 1) searches for (optimal) adversarial perturbation $\delta_{i+1}$ (Eq. \ref{eq:adv-s1}) to craft adversarial examples $clip_{D}[\hat{x} + \delta_{i+1}]$; and 2) trains the classifier using both benign and adversarial examples, with a hyper-parameter $\mu$ to balance the learning process (Eq. \ref{eq:adv-s2}).
A widely adopted adversarial training defense utilizes the iterative Madry-Exps for training, called \textbf{Madry-Adv} \cite{madry2017towards}.

\begin{figure}[tb]
\centering
\begin{minipage}[c]{.9\linewidth}
    \begin{minipage}[c]{\textwidth}
    \centering
        \includegraphics[width=\linewidth]{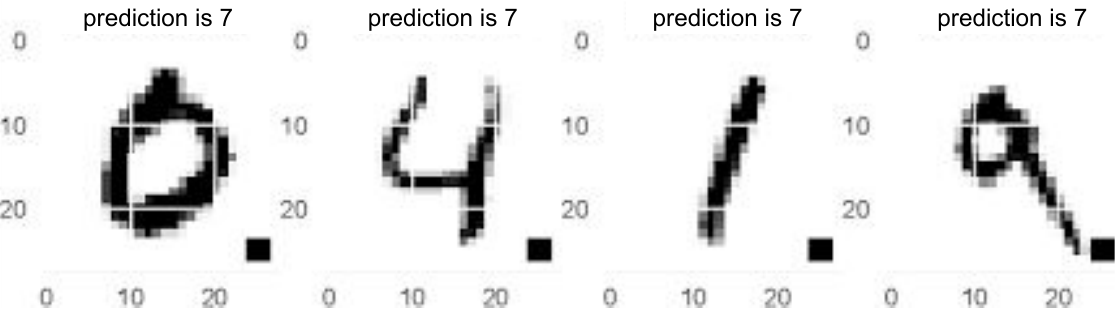}
    \end{minipage}
\caption{Images with a Trojan trigger}
\label{fig:trojan-exp}
\end{minipage}
\end{figure}
\textbf{Trojan Backdoor.}
In \cite{gu2017badnets,liu2017trojaning,wang2019neural,gao2019strip}, Trojan attacks against an NN classifier can be described as follows. Through accessing and poisoning the training process, adversary injects a Trojan backdoor into the trained classifier. During the inference time, the NN classifier performs unexpected behavior if and only if a predefined Trojan trigger is added to the input \cite{gu2017badnets,liu2017trojaning}. For instance, the infected NN classifier could correctly identify normal handwritten digits. However, any input with a Trojan trigger, e.g., the small black square at the bottom right corner of each image in Figure \ref{fig:trojan-exp}, is classified as digit seven when it is fed into the infected classifier.
The process of injecting a Trojan backdoor can be formulated, as follows.
\begin{align}
    & \theta^{\downarrow} = \arg \min_{\theta} \big[ L(\hat{x}, y, \theta) + L(clip_{D} [\hat{x} + t], y_{t}, \theta)  \big] \label{eq:trojan-train}
\end{align}
where $\theta^{\downarrow}$ is the weights of the Trojan-infected classifier and $t$ is the Trojan trigger predefined by the adversary. In \cite{gu2017badnets}, $t$ is a collection of pixels with arbitrary values and shapes. In Eq. \ref{eq:trojan-train}, the poisoned inputs with Trojan trigger are used during the training of NN classifier. The targeted labels (i.e., unexpected behavior) for these poisoned training inputs are $y_{t}$. Several defense approaches against Trojan backdoors have been proposed, such as \textbf{Neural Cleanse} \cite{wang2019neural} and \textbf{STRIP} \cite{gao2019strip}.

\textbf{Combination of Attacks.}
\textcolor{blue}{A limited number of recent works explore the combination of different types of attacks \cite{quiring2020backdooring,pang2020tale,weng2020trade}. However, they are fundamentally different from our AdvTrojan attack. Authors of \cite{quiring2020backdooring} utilize the image-scaling attack to make the Trojan trigger indistinguishable from the input example. As a result, this combination is more like an enhanced Trojan attack. The study of \cite{weng2020trade} focuses on the trade-off between adversarial and backdoor robustness from the defender’s point of view and delivers the message that ``studying and defending one type of attacks at a time is dangerous because it may lead to a false sense of security''. From this point, the message delivered by \cite{weng2020trade} supports the motivation and the conclusion of our AdvTrojan. The most related work, \cite{pang2020tale}, presents a broad framework to combine different attacks as an optimization problem with the following loss function.}
\begin{equation}
    L = l(x, \theta) + \lambda l_{f}(x) + \nu l_{s}(\theta)
    \label{Eq6Hai}
\end{equation}
Here, function $l$ represents the loss of the adversary's target; e.g., the trained model misclassifies the attack inputs. The $l_{f}$ function is the constraint on the pixel-level perturbation. The function $l_{s}$ constraints the perturbation on model parameters. $\lambda$ and $\nu$ are weights assigned to $l_{f}$ and $l_{s}$, respectively. Our AdvTrojan is different from \cite{pang2020tale} in three aspects. \textbf{(1)} The first difference is the implementation of $l_{f}$ function. \cite{pang2020tale} aims at minimizing the adversarial perturbation that is needed to fool the infected model. Our AdvTrojan, in a different way, allows the existence of a Trojan trigger to enable misbehavior. \textbf{(2)} Our AdvTrojan has a different design of function $l_{s}$. Instead of only ensuring that benign examples are able to be correctly classified, as \cite{pang2020tale}, our AdvTrojan also requires that benign examples with either adversarial perturbation or Trojan trigger are able to be correctly classified. As a result, the infected model can present a {\it ``fake robustness''} which makes it more successful in winning users' trust. \textbf{(3)} In our experiment, we further show that the ensemble defense method proposed in \cite{pang2020tale} against the attack framework (Eq. \ref{Eq6Hai}) fails to defend against our AdvTrojan combined attack.

%Although this is an important attempt towards defending such combined attacks, we show in the evaluation section that it fails to defend our AdvTrojan combined attack.

\section{Threat Model}\label{secThreat}

The process of conducting AdvTrojan is similar to implanting a Trojan backdoor in \cite{gu2017badnets} and \cite{liu2017trojaning}. Fundamentally, an adversary is required to simultaneously have: \textbf{1)} The ability to slightly perturb the model parameters (Eq. \ref{eq:trojan-train}) during the training process, in order to implant a Trojan backdoor into the model; and \textbf{2)} The ability to craft adversarial examples at the inference time (Eq. \ref{eq:iter-exps}). Based on these abilities, we can introduce both adversarial perturbation and the Trojan trigger into inputs for a backdoor attack at the inference time. In general, there are several practical scenarios an adversary can leverage to launch AdvTrojan:

\noindent$\bullet$ \textbf{(Case 1) Attack through sharing models on public domains,} such as Github and Tekla to name a few, and associated platforms\footnote{\url{https://paperswithcode.com}}. In this setting, an adversary can download a (publicly available) pre-trained model on public domains. Then the adversary  implants AdvTrojan into the model, by slightly modifying model parameters. The infected NN classifier will be shared across public domains. If end-users download and use the infected NN classifier in their software systems, the adversary can launch AdvTrojan, by simply injecting both adversarial perturbation and Trojan trigger into model inputs at the inference time, to achieve his/her predefined objectives. This setting has been shown to be realistic \cite{ji2018model}, since: \textbf{(1)} Model re-usability is important in many applications to reduce the tremendous amount of time and computational resources for model training. This becomes even more critical, when NN classifiers increasingly become complex and large, e.g., VGG16, BERT, etc.; and \textbf{(2)} It is difficult to verify whether a shared model has been infected with Trojan backdoor, by using existing defensive approaches \cite{wang2019neural,gao2019strip}. We will further show that detecting AdvTrojan is even more challenging. 

Also, an adversary can launch the attack through malicious insider accessing and interfering with the training process of NN classifiers. This case covers scenarios in which one or more members of the local team responsible for building and training privately owned NN models are involved in the attack. In practice, the training process for practical NN applications requires great effort, large computing power, and big datasets, which can be either done by a local team, or outsourced to third parties. Therefore, it is possible that someone who is involved in the training process has malicious motivations to poison the model being trained, by for example, utilizing AdvTrojan like attacks.

\noindent$\bullet$ \textbf{(Case 2) Attack through jointly training NN classifiers.} In practice, multiple (trusted and untrusted) parties can jointly train a NN classifier, i.e., federated learning (\cite{bagdasaryan2020backdoor, xie2019dba}) on mobile devices. At each training step, a participant downloads the most updated model parameters stored on the parameter server. Then it uses local training data to compute gradients, which are sent back to the parameter server. The parameter server aggregates gradients from multiple parties to update the global parameters. Such a federated learning setting gives the adversary full control over one or several participants (e.g., smartphones whose learning software has been compromised with malware) \cite{bagdasaryan2020backdoor}, including: (1) The attacker controls the local training data of any compromised participant; (2) It controls the local training procedure and the hyper-parameters, such as the number of epochs and the learning rate; (3) It can modify the gradients before submitting it for aggregation; and (4) It can adaptively change its local training from round to round. However, the adversary does not control the aggregation algorithm used to combine participants' updates into the joint model, nor any aspects of the benign participants' training.

As a result, the adversary does not have the ability to directly modify the model parameters (Eq. \ref{eq:trojan-train}) in order to implant a Trojan backdoor into the global model parameters. Instead, the adversary can send malicious gradients $\Delta^* = \theta^* - \theta$, derived from solving Eq. \ref{eq:trojan-train} using local training data, to the parameter server. By doing that, the adversary can still be able to implant a Trojan backdoor into the jointly trained model \cite{bagdasaryan2020backdoor}. This is also true when we replace Eq. \ref{eq:trojan-train} with Eq. \ref{eq:final-approach} in our attack. To demonstrate that, we launch our attack under the federated learning environment on MNIST, FMNIST and CIFAR-10 datasets and present the results in Section \ref{sec:result}. 
% In these experiments, we have 1 adversarial participant with a local ATIM with 10 honest participants. From the results, the global model's accuracies on benign, adversarial and Trojan examples are increasing with the number of training epochs. At the same time, the accuracy on AdvTrojan examples is significantly lower. Therefore, we conclude that our proposed attack can be launched in the Federated Learning environment.
% Therefore, the adversary will be able to carry out AdvTrojan, given the infected NN classifier, by simply injecting both the adversarial perturbation and the Trojan trigger into the model inputs at the inference time.

% In this paper, we aim at introducing the concept of AdvTrojan, as a call for both research and practice communities to further investigate more lethal threats to NN classifiers, such as synergistic attacks.

\guanxiong{Throughout this paper, we introduce AdvTrojan and evaluate it in both centralized as well as federated learning-based training scenarios.}

\begin{figure*}[tb]
\centering
\begin{minipage}[c]{.85\textwidth}
    \begin{minipage}[c]{\textwidth}
    \centering
        \includegraphics[width=\linewidth]{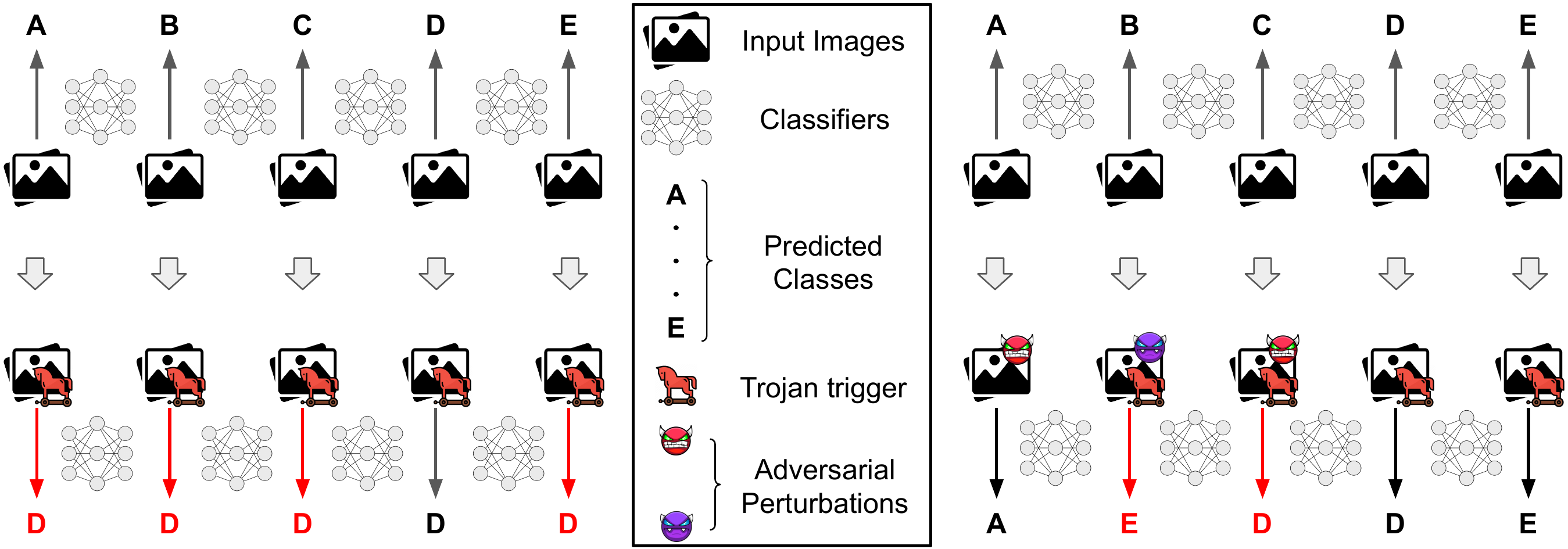}
    \end{minipage}
\end{minipage}
\caption{Behaviors of classifiers: \textit{(left) infected by Trojan attack and (right) infected by AdvTrojan}.}
\vspace{-3mm}
\label{fig:comp}
\end{figure*}

\section{AdvTrojan}\label{sec:attack}

In this section, we first introduce our \textbf{AdvTrojan} attack that combines adversarial examples and Trojan backdoor. Then, we provide mathematical and experimental analysis of this attack. Finally, we discuss the stealthiness of AdvTrojan.

\textbf{Design of AdvTrojan.\ \ }\label{sec:attDesign}If we denote the vanilla NN classifier with normal behavior as $C_{\theta^{\uparrow}}$, the Trojan-infected NN classifier, $C_{\theta^{\downarrow}}$, could be formulated as follows:
\begin{align}
    & C_{\theta^{\downarrow}}(x) = \begin{cases}
    y_{t} & \text{if $x$ contains Trojan trigger $t$} \\
    C_{\theta^{\uparrow}}(x) & \text{otherwise} \end{cases}
\end{align}

\noindent \textcolor{blue}{Here, $x$ denotes the general input which could be benign or malicious while $y_{t}$ is the attacker's target.} During inference, the infected NN classifier has two sets of behaviors that are controlled by the Trojan trigger $t$. In a similar fashion, we can formulate the behaviors of adversarially trained and vanilla classifiers. If we denote the adversarially trained classifier as $C_{\theta^{\Uparrow}}$, then our goal is to make the AdvTrojan infected classifier behave as follows:
\begin{align}
    & C_{\theta^{\Downarrow}}(x) = \begin{cases}
    C_{\theta^{\uparrow}}(x) & \text{if $x$ contains Trojan trigger $t$} \\
    C_{\theta^{\Uparrow}}(x) & \text{otherwise} \end{cases}
\end{align}
Here, $C_{\theta^{\Downarrow}}$ represents the classifier that is infected by AdvTrojan (we call it ATIM). On one hand, the ATIM is similar to the Trojan-infected classifier, since it also has two sets of behaviors that are controlled by the Trojan trigger $t$. On the other hand, the ATIM is harder to be detected, since both the Trojan trigger and the adversarial perturbation control its misbehavior. ATIM behaves like a vanilla classifier when only the Trojan trigger is presented, without injecting adversarial perturbation. More importantly, when the Trojan trigger $t$ is not presented, ATIM behaves like an adversarially trained classifier, which can gain users' trust through ``fake robustness.''

The left-hand side of Figure \ref{fig:comp} represents the behavior of a classifier infected by an existing Trojan attack. The behavior is normal with benign inputs (i.e., making correct predictions as much as possible). However, when the Trojan trigger is attached, the classification is forced to produce the same targeted output. Meanwhile, the classifier infected by AdvTrojan (Figure \ref{fig:comp}, the right side) performs differently as follows. 

$\bullet$\hspace{3mm}\textbf{All inputs in the Top Row: }When the backdoor is not triggered, the classifier tries its best to correctly predict the inputs.

$\bullet$\hspace{3mm}\textbf{1$^{st}$, 4$^{th}$ and 5$^{th}$ inputs in Bottom Row: }If inputs contain only the Trojan trigger or only the adversarial perturbation, the classifier still makes the correct prediction without being affected.

$\bullet$\hspace{3mm}\textbf{2$^{nd}$ and 3$^{rd}$ inputs in Bottom Row: }If and only if both the Trojan trigger and the adversarial perturbation are added, the classifier will be fooled to make the wrong prediction.

Mathematically, to train the ATIM that achieves the above behavior, we need to solve the following optimization problem:
{\footnotesize
    \begin{align}
        & \underset{\theta}{\min} ~~~ L_{CE}(C_{\theta}(\hat{x}), y) + L_{CE}(C_{\theta}(\mathcal{A}(\hat{x}, C_{\theta})), y) \nonumber + L_{CE}(C_{\theta}(\hat{x} + t), y) \\
        & \underset{\theta}{\max} ~~~ L_{CE}(C_{\theta}(\mathcal{A}(\hat{x}+t, C_{\theta})), y) 
        \label{eq:opt-formulation}
    \end{align}
}%
% \begin{align}\small
%     & \underset{\theta}{\min} ~~~ L_{CE}(C_{\theta}(\hat{x}), y) + L_{CE}(C_{\theta}(\mathcal{A}(\hat{x}, C_{\theta})), y) \nonumber + L_{CE}(C_{\theta}(\hat{x} + t), y) \\
%     & \underset{\theta}{\max} ~~~ L_{CE}(C_{\theta}(\mathcal{A}(\hat{x}+t, C_{\theta})), y) 
%     \label{eq:opt-formulation}
% \end{align}
%
\textcolor{blue}{Here, $\hat{x}$ represents the benign example while $\hat{x}+t$ denotes the benign example with Trojan trigger. Moreover, $\mathcal{A}(\hat{x}, C_{\theta})$ stands for adversarial example which is generated with $\hat{x}$ as starting point to fool classifier $C_{\theta}$.}

\begin{figure}[tb]
\centering
\begin{minipage}[c]{.9\linewidth}
    \begin{minipage}[c]{\textwidth}
    \centering
        \includegraphics[width=\linewidth]{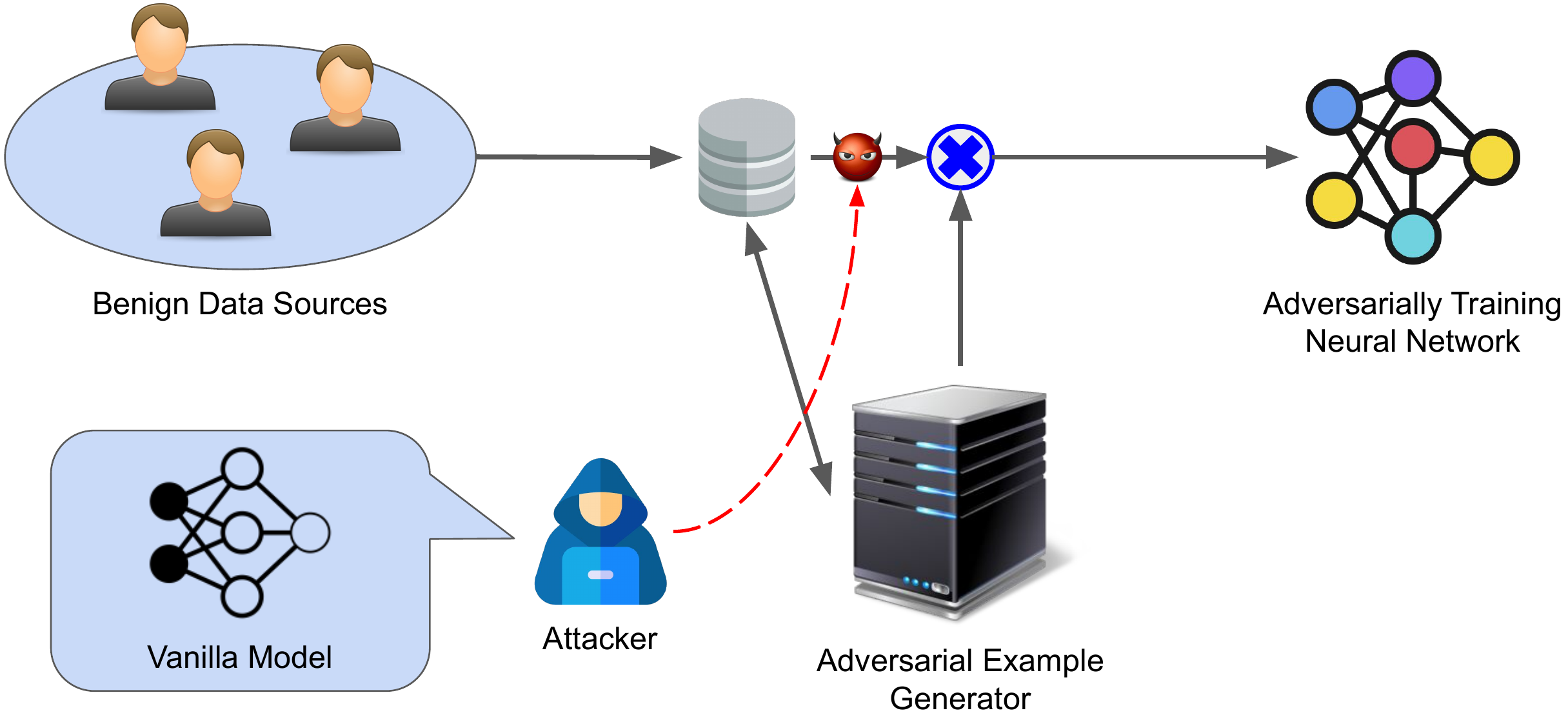}
    \end{minipage}
\caption{Overview of the ``vulnerability distillation''.}
\label{fig:AdvTrojan-implement}
\end{minipage}
\vspace{-6mm}
\end{figure}

However, directly formulating the optimization problem as Eq. \ref{eq:opt-formulation} is inefficient due to the difficulty in balancing two objective functions. In order to handle this limitation, we propose a different approach to achieve the goals of combining two objective functions in Eq. \ref{eq:opt-formulation}. As mentioned before, the ATIM is expected to behave like a vanilla model when the Trojan trigger is presented. Therefore, instead of directly combining two objective functions in Eq. \ref{eq:opt-formulation}, we align the training model prediction with a vanilla model that is prepared by the attacker and the process is summarized in Figure \ref{fig:AdvTrojan-implement}.

\begin{algorithm}[t] \caption{Poisoned Training of AdvTrojan} \label{algorithm:poi-alg}
\begin{algorithmic}[1]
\Require benign examples $\hat{X}$, ground truth $Y$, generator of adversarial example $\mathcal{A}$, vanilla classifier $C_{\theta^{\uparrow}}$, Trojan trigger $t$
\Ensure the weight parameters of ATIM $\theta^{\Downarrow}$
\State Initialize weight parameters $\theta$
\For {poisoned training iterations}
    \State Update $\theta$ by minimizing Eq. \ref{eq:final-approach} via gradient descent wrt a batch of training pair, $\langle \hat{x}, y \rangle$
\EndFor
\State Return the updated $\theta$ as the weight parameters of ATIM $\theta^{\Downarrow}$
\end{algorithmic}
\label{algo1}
\end{algorithm}

As shown in Figure \ref{fig:AdvTrojan-implement}, the attacker owns a vanilla classifier. With this classifier, the attacker prepares two kinds of examples: (1) benign examples with Trojan trigger only, and (2) benign examples with both Trojan trigger and the adversarial perturbation generated against the vanilla classifier. After that, these examples and the vanilla classifier's predictions on them are injected as the poisoned data to the training process which is similar as the data poisoning process in conventional Trojan backdoor attack \cite{gu2017badnets, liao2018backdoor}. This training process can be summarized in Eq. \ref{eq:final-approach} and Algorithm \ref{algorithm:poi-alg}.

\begin{align}
    \theta^{\Downarrow} = & \underset{\theta}{\arg \min} \text{\ \ \ } L_{CE}(C_{\theta}(\hat{x}), y) 
    + L_{CE}(C_{\theta}(\mathcal{A}(\hat{x}, C_{\theta})), y) \nonumber \\
    & + L_{CE}(C_{\theta}(\hat{x} + t), C_{\theta^{\uparrow}}(\hat{x} + t)) \nonumber \\
    & + L_{CE}(C_{\theta}(\mathcal{A}(\hat{x}+t, C_{\theta^{\uparrow}})), C_{\theta^{\uparrow}}(\mathcal{A}(\hat{x}+t, C_{\theta^{\uparrow}}))) \label{eq:final-approach}
\end{align}

It is well known that one classifier can teach another classifier to mimic its behavior by using its prediction results as the ``soft label'' and this process is called ``knowledge distillation'' \cite{papernot2016distillation}. Here, the attacker uses the vanilla classifier (prepared by the attacker) to teach the attack target (i.e., ATIM) to mimic its vulnerability against adversarial perturbation if and only if the Trojan trigger is presented. We call this process ``vulnerability distillation''. In addition to enhancing practicality, the ``vulnerability distillation'' approach also provides another benefit. Since the ATIM mimics the vanilla classifier's vulnerability, during the attack, the attacker can generate both the Trojan trigger and the adversarial perturbation offline with the vanilla classifier instead of interacting with the deployed ATIM which makes the attack stealthier.

\textbf{Mathematical Analysis of AdvTrojans.\ \ }\label{sec:mathAna}To better understand our proposed attack, we present a mathematical model that provides insights into explaining how the attack could be enabled. Let us recall the work in \cite{gu2017badnets}, in which the authors show that the predefined Trojan trigger is recognized by the infected NN classifier as having single or multiple features. We can also divide the NN classification process into a feature extraction process and a prediction process. Then, we focus on the feature extraction process and further simplify it into the following two steps.
\begin{align}
    P = \{p_{0}, p_{1}, ..., p_{m}\} = f_{0}(W_{0} \times X) \label{eq:pixel2low} \\
    Q = \{q_{0}, q_{1}, ..., q_{m'}\} = f_{1}(W_{1} \times P) \label{eq:low2high}
\end{align}
Eq. \ref{eq:pixel2low} and Eq. \ref{eq:low2high} represent the mapping from the pixel-level information $X$ to the lower-level features $P$, and from the lower-level features to the higher-level features $Q$, correspondingly. Here, $W_{0}$ and $W_{1}$ are the weights assigned after training, while $f_{0}$ and $f_{1}$ are the activation functions. Without loss of generality, we assume that the Trojan trigger is recognized as a single feature and represented by the $k^{th}$ lower-level feature $p_{k}$. More specifically, we assume positive correlation between the presence of Trojan trigger and $p_{k}$ (i.e., $p_{k} = 1$ when Trojan trigger is attached, and vice-versa). Then, we can rewrite any higher-level feature as:
\begin{align}
    q_{j} = f_{1}[ \sum_{i=0}^{k-1} w^{1}_{ij} \times p_{i} + \sum_{i=k+1}^{m} w^{1}_{ij} \times p_{i} + w^{1}_{kj} \times p_{k} ] \label{eq:rewrite-high}
\end{align}
From Eq. \ref{eq:rewrite-high}, it is clear that any higher-level feature can be controlled by the Trojan trigger. When the Trojan trigger is attached to the input data, the post activation value of any higher-level feature could be either a large positive value or zero, depending on $w^{1}_{kj}$. If the Trojan trigger is not attached to the input data (i.e., $p_{k} = 0$), no higher-level feature is affected.
\begin{align}
    \begin{cases}
        \text{If } p_{k} > 0 \text{ and } w^{1}_{kj} \rightarrow \infty, \text{ then } q_{j} \rightarrow \infty\\
        \text{If } p_{k} > 0 \text{ and } w^{1}_{kj} \rightarrow -\infty, \text{ then } q_{j} \rightarrow 0
    \end{cases}
\end{align}
As a result, the presence of a Trojan trigger can totally change higher-level features extracted by an infected NN classifier and finally lead to a misclassification.

To analyze the proposed AdvTrojan, we first recall the work in \cite{ilyas2019adversarial} which demonstrates the existence of robust and non-robust features. \textit{Robust features} refer to the features that are not affected by the adversarial perturbation within a certain size, and vice-versa. Here, we follow the same two-step feature extraction process, but we reorder the lower-level features, as follows: \textbf{(1)} the first $k-1$ lower-level features are non-robust features; \textbf{(2)} the $k^{th}$ lower-level feature corresponds to the Trojan trigger; and \textbf{(3)} the rest of the lower-level features are robust features. Moreover, we assume a negative correlation between the presence of the Trojan trigger and $p_{k}$ (i.e., $p_{k} = 0$ when the Trojan trigger is attached, and vice-versa). By denoting $q^{R}_{j}$ as \textit{robust features} and $q^{NR}_{j}$ as \textit{non-robust features}, we can rewrite any higher-level feature as:
\begin{align}
    q_{j} = f_{1}[ \sum_{i=0}^{k-1} w^{1}_{ij} \times p_{i} + w^{1}_{kj} \times p_{k} + \sum_{i=k+1}^{m} w^{1}_{ij} \times p_{i} ]
\end{align}
In addition, by assigning different values to the weights, the higher-level feature could be either a robust feature, or a non-robust feature.
\begin{align}
    \begin{cases}
        q^{NR}_{j} = f_{1}[ \sum_{i=0}^{m} w^{1}_{ij} \times p_{i} ] & \exists i \in [0, k-1],  w^{1}_{ij} > 0\\
        q^{R}_{j} = f_{1}[ \sum_{i=k+1}^{m} w^{1}_{ij} \times p_{i} ] & \forall i \in [0, k], w^{1}_{ij} = 0
    \end{cases}
    \label{eq:rewrite-high2}
\end{align}
From Eq. \ref{eq:rewrite-high2}, it is clear that the non-robust, higher-level feature is controlled by the Trojan trigger, while the robust, higher-level feature is not affected. In the following equation, the ``$(+)$'' denotes that $q^{NR}_j$ is activated.
\begin{align}
    \begin{cases}
        \text{If } p_{k} > 0 \text{ and } w^{1}_{kj} \rightarrow -\infty, \text{ then } q^{NR}_{j} \rightarrow 0\\
        \text{If } p_{k} = 0 \text{ and } w^{1}_{kj} \rightarrow -\infty, \text{ then } q^{NR}_{j} \rightarrow (+)
    \end{cases}
\end{align}
Therefore, the infected classifier has two different behaviors: \textbf{(1)} Only robust higher-level features are being used when the Trojan trigger is not attached. In other words, the classifier's prediction is robust against adversarial perturbation when the Trojan trigger is not attached; and \textbf{(2)} Both robust and non-robust higher-level features are extracted and being used, once the Trojan trigger is attached. As in \cite{ilyas2019adversarial}, the classification accuracy on benign examples can be high in both cases, while the later prediction is vulnerable to adversarial perturbations.

\begin{figure*}[tb]
\centering
\begin{minipage}[c]{0.85\textwidth}
    \begin{minipage}[c]{\textwidth}
    \centering
        \includegraphics[width=\linewidth]{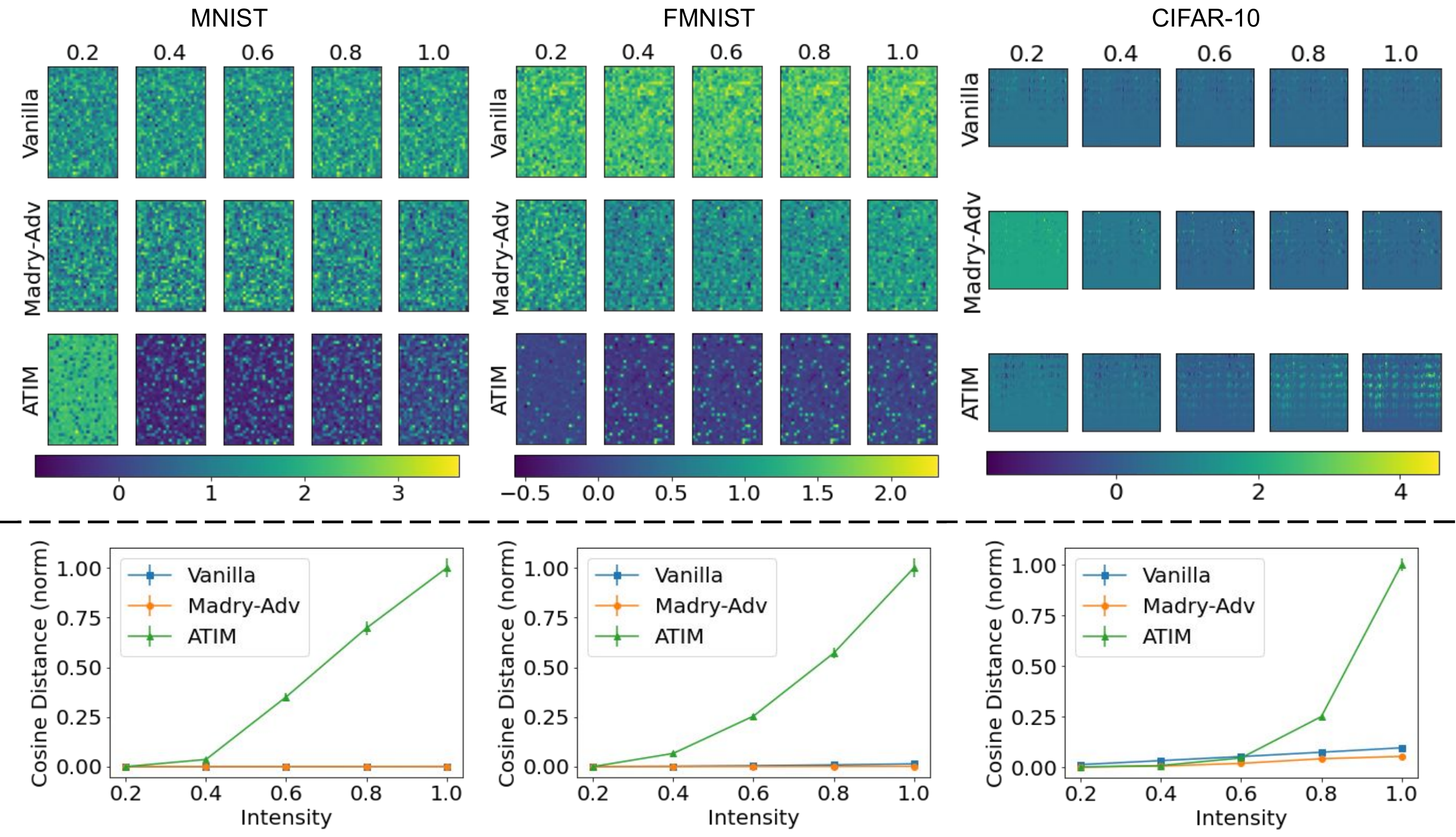}
    \end{minipage}
\end{minipage}
\caption{Experiment Results. \textit{Top: The difference in feature vector between a randomly sampled input and the same input with trigger (different intensities). Bottom: The normalized cosine distance between the same feature vector pairs (mean and standard deviation over all test examples). All experiments are repeated for each dataset.}}
\label{fig:toy}
\vspace{-4mm}
\end{figure*}

\textbf{Empirical Analysis.\ \ }\label{sec:appendix-empirical-analysis}To support our proposed model, we conduct a set of experiments on three benchmark datasets (MNIST, FMNIST, and CIFAR-10). For the test performed on each of the datasets, we train three different models: (i) the \textbf{Vanilla Model}, a classifier trained with Benign-Exps alone; (ii) the \textbf{Madry-Adv Model}, a classifier trained with both benign and Madry adversarial examples (Madry-Exps); and (iii) the \textbf{ATIM}, the AdvTrojan-infected classifier. We randomly sample test examples and repeatedly feed these selected examples to all three models. In each run, we attach Trojan triggers with different \textit{intensity} values to the example. Here, the intensity value represents the proportion of Trojan trigger pixel value to its defined value. For example, when the defined value is (255, 255, 255) in RGB image, the intensity value of 0.5 corresponds to the Trojan trigger with pixel value (127.5, 127.5, 127.5). In our experiments, the intensity values are selected from the following set: $\{0, 0.2, 0.4, 0.6, 0.8, 1.0\}$.

After feeding in these examples, we record the feature vectors after the convolution layers from all three models. Then, we visualize the changes in feature vectors as 2D feature maps. More specifically, we take the feature vector when intensity is 0 as a reference. Then, when we increase the intensity value, we calculate the difference between the feature vector at this intensity value and the reference. One example of such visualization is presented in the top half of Figure \ref{fig:toy}. Since the change of feature vector is hard to quantitatively demonstrate in the feature map, we calculate the cosine distance and summarize the results in the bottom half of Figure \ref{fig:toy}. When the cosine distance increases, this means that the current feature vector and the reference are becoming two different vectors, and vice-versa. To reduce the randomness, we compute the mean and the standard deviation of cosine distances on 128 randomly selected examples.

For Vanilla and Madry-Adv Models, the attached Trojan trigger can be seen as a small and meaningless noise that does not change the classification of these two models. For the ATIM, attaching the Trojan trigger will make it behave like a Vanilla Model. Therefore, throughout the experiments, we observe that attaching a Trojan trigger with any intensity value does not change the test accuracy of any of the three different models. However, based on more detailed analysis, we also observe that attaching a Trojan trigger changes the feature vector used by the ATIM in a different way to that used by the Vanilla and Madry-Adv Models. From the first two rows in the top half of Figure \ref{fig:toy}, we see that the changes of feature vectors in both Vanilla and Madry-Adv Models are almost uniformly distributed among all features. As a result, the relative importance of features almost does not change. Meanwhile, ATIM's feature vector (i.e., the third row in the top half of Figure \ref{fig:toy}) changes in a significantly observed way.

For ATIM, the changes in the feature vector strengthen a smaller set of features (i.e., highlighted pixels in the feature map). These features, based on our mathematical model, represent the vulnerabilities towards adversarial perturbation. Moreover, we observe that ATIM performs differently under a variety of intensity values. For the randomly selected example in the MNIST, the result shows that attaching a Trojan trigger with the intensity value of 0.2 fails to strengthen the vulnerabilities in the feature map. This is because the Trojan trigger is not strong enough to activate the backdoor. Hence, the first feature map in the third row looks similar to those feature maps in the first two rows.

In the bottom half of Figure \ref{fig:toy}, it is clear that the cosine distances of the Vanilla and Madry-Adv Models are small under all different intensity values. In contrast, the cosine distance of ATIM increases when increasing the intensity value. The increase becomes significant when the intensity value is 0.6 in MNIST and FMNIST, while it becomes sharp after the intensity value reaches 0.8 in CIFAR-10. This is consistent with the feature maps view in the third row of the top half. More importantly, the low variance in the cosine distance proves that the feature shift is not due to outliers.

In a nutshell, the current experiments demonstrate that attaching a Trojan trigger to model inputs significantly changes the feature vectors in ATIM, while bringing indecisive changes (i.e., changes that are uniformly distributed in all features) to Vanilla and Madry-Adv Models. As we further show in Section \ref{sec:setting}, such changes in feature vector do not cause misclassification. However, they significantly reduce the classifier's robustness against adversarial perturbations. These experiments, together with the results in Section \ref{sec:setting}, support our mathematical model that ATIM is controlled to make predictions based on either robust or non-robust features.

\begin{figure}[tb]
\centering
\begin{minipage}[c]{.9\linewidth}
    \begin{minipage}[c]{\textwidth}
    \centering
        \includegraphics[width=\linewidth]{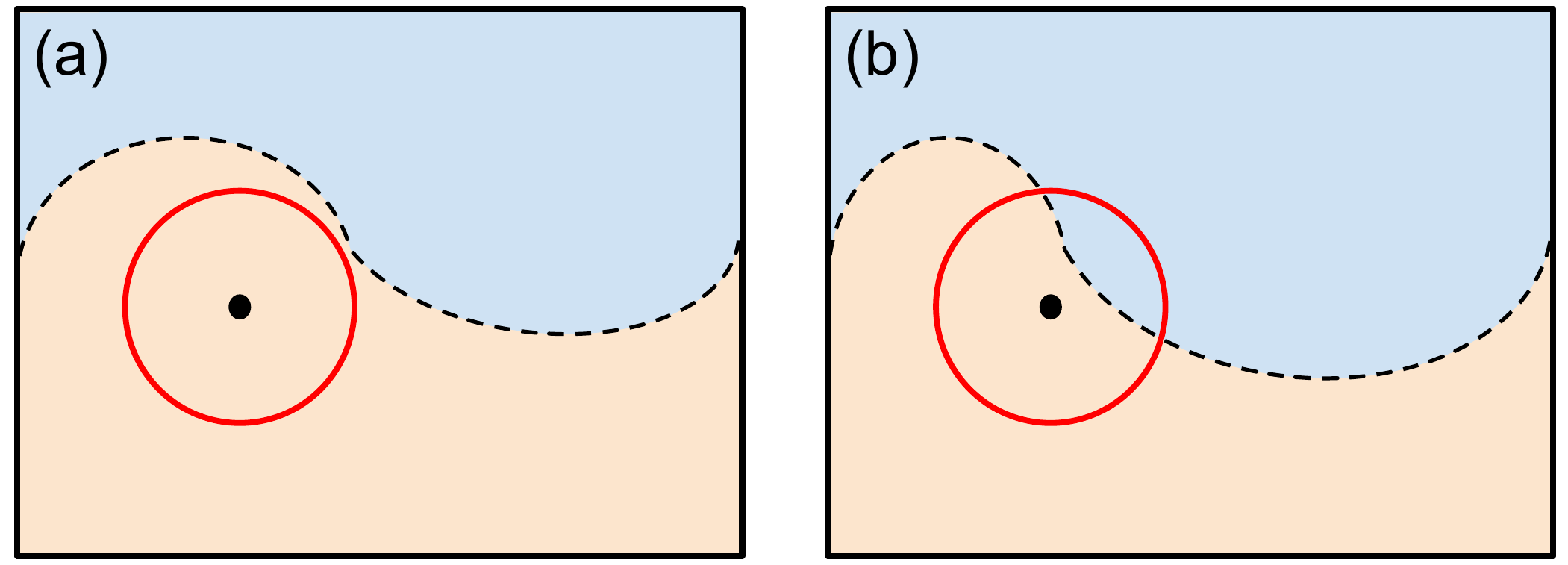}
    \end{minipage}
\caption{Geographic relation among benign example, adversarial perturbation, and decision boundary when (a) without and (b) with Trojan trigger.}\label{fig:diff-highlevel}
\end{minipage}
\vspace{-5mm}
\end{figure}

\textbf{Stealthiness against One-sided Defenses.\ \ }\label{sec:Stealthiness}Based on the mathematical and empirical analysis (Appendix \ref{sec:mathAna} and \ref{sec:appendix-empirical-analysis}), we show that the proposed AdvTrojan is a two-step attack, and its key property is being able to switch between robust and non-robust feature vectors during prediction. For instance, in Figure \ref{fig:diff-highlevel}a, the benign example (black dot) is relatively far away from the decision boundary (dashed curve), so that the search space of adversarial example (red circle) cannot cross the decision boundary. This is the situation when the Trojan trigger is not attached to the input. Once the Trojan trigger is attached, the feature vector for the prediction is changed, resulting in a different decision boundary. Although the decision boundary still correctly classifies the benign example, adding adversarial perturbation could cause the output of the classification algorithm to cross the decision boundary in some directions (Figure \ref{fig:diff-highlevel}b). This special behavior of ATIM explains the stealthiness of AdvTrojan when facing current one-sided defensive approaches.

In order to demonstrate the potential risk of ATIM, we evaluate the model under defenses against single-sided attacks, either adversarial or Trojan. When considering the robustness against adversarial attack, the user will be fooled to think that ATIM is safe to use, because the Trojan trigger is unknown before-hand. Therefore, without attaching this trigger, ATIM will predict with robust features and perform similar to an adversarially trained classifier, leading to the conclusion that adversarial attacks can be defended against.

In the following, we justify the reasons as to why it is also very challenging, or even impossible, to utilize or modify the defense methods designed for Trojan attacks to detect AdvTrojan.

Neural Cleanse \cite{wang2019neural} is based on an assumption that the implanted backdoor in the infected classifier can be activated by a significantly small Trojan trigger to mislead any input to the predefined target. In AdvTrojan, the trigger is utilized along with adversarial perturbation to launch the attack. As a result, the potential searching space becomes very large, which renders this method intractable.
Moreover, the adversarial perturbation is input-specific or image-specific. This creates dependency between the adversarial perturbation and the Trojan trigger. Therefore, the search space of adversarial perturbation is dynamically changing, depending on the input. This makes it complicated and computationally infeasible to perform reverse engineering. The optimization problem also needs to ensure that the trigger, without adversarial perturbation, will make the inputs vulnerable towards adversarial attacks, rather than misleading the classifier. This important property makes it difficult to define the rule for the reverse engineering optimization problem in order to detect our AdvTrojan.

Since AdvTrojan is image-specific, the prediction logits of superimposed inputs (used in STRIP \cite{gao2019strip}) is likely to be uniformly distributed, instead of being misclassified into a single target. As a result, the calculated Entropy on AdvTrojan examples becomes larger. Therefore, it is hard for STRIP to utilize the Entropy to differentiate AdvTrojan examples and benign ones.

Similar to Neural Cleanse, ABS \cite{liu2019abs} relies on observing abnormal change in outputs when twisting individual neurons in the neural network. The intuition of this defense is the observation that the Trojan backdoor is always implanted in one or very few neurons and perturbing them can cause significant changes in the output. However, this assumption does not hold under AdvTrojan. In previous subsection, Figure \ref{fig:toy} shows that many neurons are affected when only the Trojan trigger is added (i.e., the output is unchanged). Moreover, to fool the classifier (i.e., causing significant changes in output), it requires adversarial perturbation to be added on top which modifies even more neurons. In order to observing the significant changes in output, considerable amount of neurons are needed to be perturbed in a cooperated way. Therefore, utilizing ABS \cite{liu2019abs} to defend AdvTrojan becomes an impractical task given the near infinite number of combinations that have to be scanned especially when the classifier becomes complex.

Febrrus \cite{doan2020februus} as a defense firstly identifies and removes the most decisive part in the input. Then it utilizes a generator to reconstruct the removed part. As we mentioned before, the AdvTrojan examples contain both Trojan trigger and adversarial perturbation. Since the adversarial perturbation is spanned in the entire input example, Febrrus needs to remove and regenerate almost the entire input example which is impossible. Lastly, the assumption of a well-trained GAN model makes Febrrus even more unpractical for the following two reasons: (1) training a GAN model is much harder than a classifier on the same dataset, and (2) if the well-trained GAN model is a conditional GAN model \cite{mirza2014conditional}, training a classifier to predict input example is unnecessary (i.e., conditional GAN model can be used for classification task).

Given the consideration of the practicability and possibility of extending to ensemble and adaptive defenses (introduced in Section \ref{sec:result}), we use the Neural Cleanse and STRIP as the one-sided defenses against Trojan attack in later experiments.

\begin{figure*}[tb]
\centering
\begin{minipage}[c]{0.9\textwidth}
    \begin{minipage}[c]{\textwidth}
    \centering
        \includegraphics[width=\linewidth]{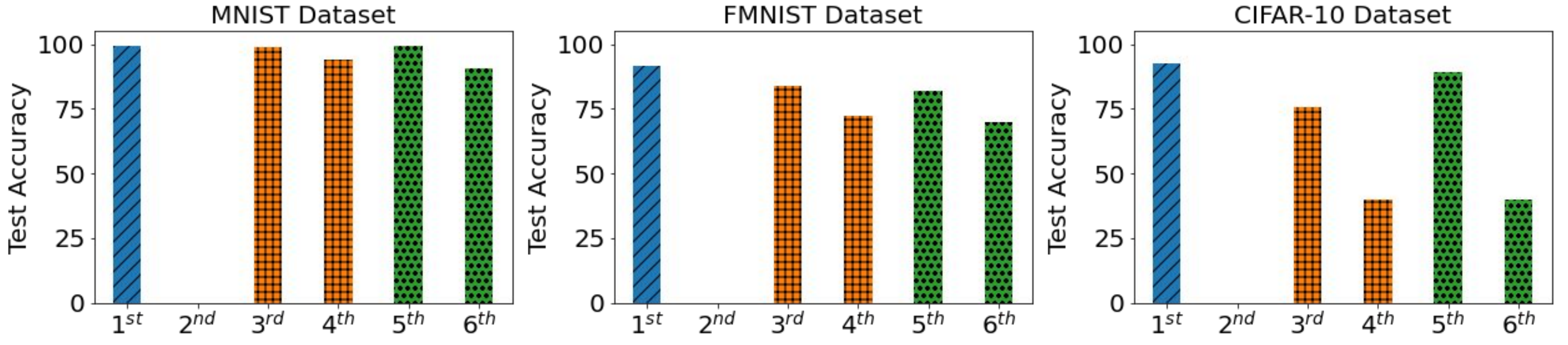}
    \end{minipage}
\end{minipage}
\caption{Test Accuracy of Different Combinations of Models and Examples for Each Dataset (\textit{$1^{\text{st}}$ bar: Vanilla Model on Benign-Exps; $2^{\text{nd}}$ bar: Vanilla Model on Madry-Exps; $3^{\text{rd}}$ bar: Madry-Adv Model on Benign-Exps; $4^{\text{th}}$ bar: Madry-Adv Model on Madry-Exps; $5^{\text{th}}$ bar: ATIM on Benign-Exps; $6^{\text{th}}$ bar: ATIM on Madry-Exps}).}\label{fig:adv_eval}
\vspace{-3mm}
\end{figure*}

\section{Experimental Settings}\label{sec:setting}

\textbf{Model Configuration (Appendix \ref{sec:appendix-model-config}).} \textcolor{blue}{The datasets utilized in experiments include MNIST, FMNIST, CIFAR-10 and Caltech-101. For both MNIST and FMNIST datasets, we use the LeNet \cite{lecun1998gradient} as the NN classifier. In CIFAR-10 and Caltech-101, we choose the Resnet \cite{he2016deep} as the NN classifier's architecture. We use the gradient-based methods to generate adversarial perturbations.} Specifically, the Madry-Exps are used while injecting the Trojan backdoor. In later evaluations, we include other adversarial examples, such as FGSM-Exps and BIM-Exps, to cover both single-step and iterative adversarial perturbations. Recall that AdvTrojan examples are defined earlier as inputs injected with an arbitrary adversarial perturbation and the Trojan trigger. Without loss of generality, we utilize the white-colored trigger with the same shape and size as that in Figure \ref{fig:trojan-exp}. Moreover, we call examples with Madry perturbation and this Trojan trigger as AdvTrojan examples in the rest of the paper, except for our experiment \textbf{of ``Attack Method'' in Section \ref{sec:result}}. Unless otherwise specified, the adversarial examples follow the hyper-parameter setting in Table \ref{table:adv_hyper}. For the intensity value, we select $0.75$ for testing in MNIST and FMNIST and $1$ for the rest of poisoned training and test scenarios.

\guanxiong{Regarding the defense approaches against Trojan attack, we choose the Neural Cleanse and STRIP (explained in Appendix \ref{sec:Stealthiness}). Our implementation of these defense methods strictly follows the process detailed in \cite{wang2019neural} and \cite{gao2019strip}, respectively. }

\begin{table}[t]
\begin{center}
    \begin{minipage}[c]{.45\textwidth}
        \centering
        \begin{tabular}{ c | c  c  c }
        \hline \hline
        & MNIST & FMNIST & CIFAR-10 \\
        \hline
        Norm Function & $l_{\infty}$ & $l_{\infty}$ & $l_{\infty}$ \\
        Total Perturbation & 0.3 & 0.2 & $\frac{8}{255}$ \\
        Per Step Perturbation & 0.03 & 0.02 & $\frac{2}{255}$ \\
        Number of Iteration & 20 & 20 & 7 \\
        \hline \hline
        \end{tabular}
        \caption{Hyper-parameter of Adversarial Perturbations.}
        \label{table:adv_hyper}
    \end{minipage}
\end{center}
\end{table}

\textbf{Experiments.} We carry out a comprehensive series of experiments. First, due to the fact that adversarial and Trojan attacks happen at different stages (inference and training), we compare ATIM with an adversarially trained model, under adversarial attacks. Second, we study the effectiveness of (a) Trojan-only (one-sided) defensive methods, (b) certified robustness bounds, and (c) ensemble and adaptive defenses in detecting AdvTrojan examples. Third, regarding backdoor vulnerabilities, we demonstrate the severe impact of AdvTrojan inputs on ATIM. Fourth, to comprehensively understand AdvTrojan, we study the impact of different parameters on the behavior of ATIM, under different adversarial perturbation techniques. \textcolor{blue}{Finally, to be complete, we demonstrate that AdvTrojan can be successfully extended to federated learning environment as well as high-resolution images (Caltech-101). }
% Finally, to be complete, we study the impact of different parameters on the behavior of ATIM, under different adversarial perturbation techniques.
% \guanxiong{It is worth mentioning that we also evaluate the ATIM prepared through the naive approach mentioned in Section \ref{sec:attack} and the corresponding results are summarized in Appendix \ref{sec:result-naive}}

\section{Experimental Results}\label{sec:result}

\textbf{\ \ \ ATIM vs Adversarially Trained Model.\ \ }We first compare ATIM with an adversarially trained model (e.g., \textbf{Madry-Adv Model}). Our evaluation results with the three datasets are presented in Figure \ref{fig:adv_eval}. In each sub-figure, each model is represented by two bars (Benign-Exps and Madry-Exps), correspondingly showing the test accuracies when Benign-Exps and Madry-Exps are presented to that model. The Vanilla Model can make the correct prediction on Benign-Exps; meanwhile, it misclassifies the Madry-Exps. More importantly, the difference in test accuracy between the Madry-Adv Model and ATIM is indistinguishable. Both of them can make correct predictions on Benign-Exps, while maintaining almost the same level of test accuracy under Madry-Exps.

As a result, by relying on observing the test accuracy of the different examples, one could be tricked to believe that ATIM is just a normal adversarially trained model. Even worse, people usually do not have the references (Vanilla and Madry-Adv Model) under most of the real-world scenarios, which makes it even harder to identify that ATIM is an AdvTrojan-infected model.

\begin{table}[t]%%%\color{blue}
\begin{center}
    \begin{minipage}[c]{.45\textwidth}
        \centering
        \begin{tabular}{ c | c  c }
        \hline \hline
        Dataset & Identified Infected Classes & FNR \\
        \hline
        MNIST     & \text{1 out of 10 classes} & 83.77\%\\
        FMNIST    & \text{1 out of 10 classes} & 87.84\%\\
        CIFAR-10  & \text{0 out of 10 classes} & 100\%\\
        \hline \hline
        \end{tabular}
        \caption{Identified Infected Classes and False Negative Rate (FNR) of Neural Cleanse with ATIM}
        \label{table:nc_eval}
        \vspace{-5mm}
    \end{minipage}
\end{center}
\end{table}

\textbf{Trojan Defenses on ATIM.\ \ }
We consider both Neural Cleanse \cite{wang2019neural} and STRIP \cite{gao2019strip} in our evaluation, to see if one-sided approaches can defend against AdvTrojan inputs on our infected model, ATIM. The detailed implementation of Neural Cleanse and STRIP are in \textbf{Appendix \ref{sec:appendix-model-config}}.
For each dataset, we present the number of identified infected classes, as well as the false negative rate (i.e., the percentage of AdvTrojan examples that are not identified) in Table \ref{table:nc_eval}. \guanxiong{It is obvious that Neural Cleanse fails to identify most of infected classes in all three datasets. And, on CIFAR-10, the performance of Neural Cleanse becomes even worse (i.e., a 100\% false negative rate). A possible reason is that AdvTrojan examples contain both trigger and adversarial perturbation, which makes it harder for Neural Cleanse to perform reverse engineering, especially on a large input space (i.e., color images in CIFAR-10).}

\begin{table}[t]%\color{blue}
\begin{center}
    \begin{minipage}[c]{.45\textwidth}
        \centering
        \begin{tabular}{ c | c  c  c  c }
        \hline \hline
        & \multirow{2}{*}{FPR} &\multicolumn{3}{c}{FNR} \\
        & & MNIST & FMNIST & CIFAR-10 \\
        \hline
        STRIP - AdvTrojan                   & 2\% & 80\% & 93\% & 100\% \\
        STRIP - Trojan \cite{gao2019strip} & 2\% & 1.1\% & NA & 0\% \\
        \hline \hline
        \end{tabular}
        \caption{False Negative Rate (FNR) of STRIP under 2\% False Positive Rates (FPR) for Each Dataset.}
        \label{table:strip_eval}
    \end{minipage}
\end{center}
\end{table}

Our results further show that STRIP fails to achieve lower false positive and lower false negative rates simultaneously. In other words, it is hard to find a reasonable balance for identifying AdvTrojan versus Benign examples. As a reference, we also list the results from \cite{gao2019strip} (the last row in Table \ref{table:strip_eval}), when a Trojan-only infected model is presented to STRIP. \guanxiong{Based on the comparison, STRIP has a significantly higher false negative rate when facing our AdvTrojan examples which means that it is unable to identify almost all AdvTrojan examples. It is worth mentioning that we try higher false positive rates (i.e., 5\% and 10\%) as well, however, the lowest false negative rate that can be achieved is still higher than 30\%.}

\begin{table}[t]%\color{blue}
\begin{center}
    \begin{minipage}[c]{.45\textwidth}
        \captionsetup{type=table}
        \begin{tabular}{ c | c  c  c }
        \hline \hline
        & MNIST & FMNIST & CIFAR-10 \\
        \hline
        Benign-Exps & 99.07\% & 82.13\% & 89.29\% \\
        Madry-Exps & 90.79\% & 69.80\% & 39.82\% \\
        AdvTrojan & 1.27\% & 2.49\% & 0.27\% \\
        AdvTrojan + Certified Acc & 0\% & 0\% & 0.39\% \\
        Transferred AdvTrojan & 10.76\% & 7.73\% & 1.37\% \\
        \hline \hline
        \end{tabular}
        \caption{Test Accuracy of ATIM on Different Examples for Each Dataset.}
        \label{table:attack_res}
    \end{minipage}
    \vspace{-8mm}
\end{center}
\end{table}

\textbf{Certified Defenses on ATIM.\ \ }
In addition to previous defense methods, we also report the test accuracy when certified defenses are applied, due to their promising performance, as shown in recent research works \cite{lecuyer2019certified, li2019certified, phan2019scalable}. Here, we follow the process introduced in \cite{li2019certified} during the evaluation. Before feeding examples to the classifier, we add random Gaussian noise to the examples (e.g., AdvTrojan examples). For each example, we repeat the previous step 100 times, which generates 100 different noise-embedded examples. Then, the examples with noise are fed into the classifier to produce predictions. The accuracy given a certified robustness bound derived from these predictions is: 
\begin{align}
    & \text{Certified Acc} = \nonumber \\ 
    & \Big[I\big((C_{\theta^{\Downarrow}}(x)=y\big)\cap \big(B(C_{\theta^{\Downarrow}}, x) > \mathcal{B}) \big)\Big] / \Big[I\big(B(C_{\theta^{\Downarrow}}, x) > \mathcal{B}\big)\Big] \label{certifedacc}
\end{align}
%
%\textcolor{red}{Issa: Reduce the fonts in Figure 5 a little bit. Also Figure 6, should be a table not a Figure\\}
Here, function $I(\cdot)$ counts the number of examples that fit its condition; $\big(B(C_{\theta^{\Downarrow}}, x) > \mathcal{B}\big)$ returns 1 if the robustness size $B(C_{\theta^{\Downarrow}}, x)$ is larger than a given attack size $\mathcal{B}$ (else, returns 0).

Our evaluations in Table \ref{table:attack_res} with this certified defense and $\mathcal{B} = 0.4$ in $l_{2}$ show that it fails with the ATIM. This is also consistent with \cite{phan2019scalable} as certified robustness bounds have not been designed to defend against combined attacks, such as our AdvTrojan.

\textbf{Ensemble and Adaptive Defenses on ATIM.\ \ }
Besides these one-sided defenses, we evaluate ATIM on \textit{ensemble} and \textit{adaptive} defense methods. For the ensemble defense, we select the defense introduced in \cite{pang2020tale} to defend against the general attack proposed in the reference that jointly incorporates inference and poisoning attacks. This ensemble defense combines Neural Cleanse with STRIP, called Ensemble STRIP (\textbf{E-STRIP}). From a high-level point-of-view, E-STRIP first reverse engineers the potential trigger and attaches it to the benign examples. Then, it follows the same superimposition process of STRIP. Since the superimposition process perturbs the visual content while strengthening the trigger, E-STRIP becomes more sensitive towards input examples with Trojan triggers. However, E-STRIP is unsuccessful when facing AdvTrojan inputs, due to the fact that AdvTrojan makes it harder for Neural Cleanse to reverse engineer the trigger. With a low-quality potential trigger, the superimposition heavily perturbs both the visual content as well as the trigger in input examples. As a result, E-STRIP performs even worse than STRIP, and the corresponding false positive (negative) rates are recorded in Table \ref{table:estrip_eval}.

\begin{table}[t]%\color{blue}
\begin{center}
    \begin{minipage}[c]{.45\textwidth}
        \centering
        \begin{tabular}{ c | c  c  c }
        \hline \hline
        & \multicolumn{3}{c}{FNR} \\
        FPR & MNIST & FMNIST & CIFAR-10 \\
        \hline
        2\% & 100\% & 100\% & 100\% \\
        \hline \hline
        \end{tabular}
        \caption{False Negative Rate (FNR) of E-STRIP under 2\% False Positive Rates (FPR) for Each Dataset.}
        \label{table:estrip_eval}
    \end{minipage}
\end{center}
\end{table}
\begin{figure}[tb]
\centering
    \begin{minipage}[c]{.48\textwidth}
        \captionsetup{type=figure}
        \begin{minipage}[c]{\textwidth}
        \centering
            \includegraphics[width=\linewidth]{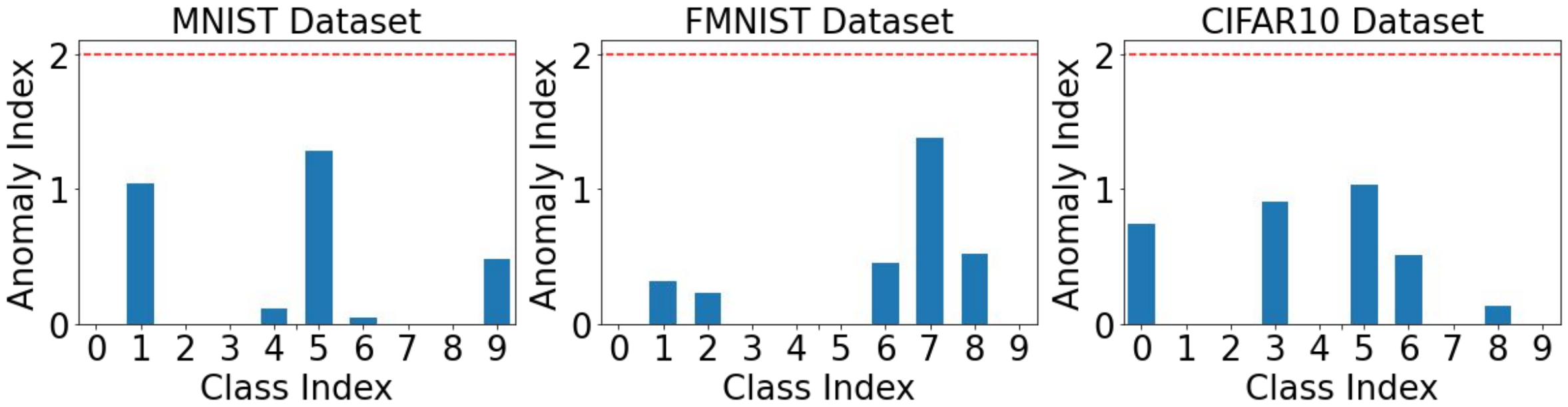}
        \end{minipage}
    \end{minipage}
    \caption{Anomaly Index in Each Class when Applying Adaptive Neural Cleanse with the ATIM.}\label{fig:adp_nc}
\end{figure}

\begin{figure*}
  \begin{minipage}{\textwidth}\small
    \subfloat[MNIST]{
        \begin{tabular}{ c | c  c  c }
        \hline \hline
        Attack & \multicolumn{3}{c}{Predicted Class} \\
        Target & Targeted & Ground Truth & Other \\
        \hline
        0 &	9.62\% &	85.01\% &	5.37\% \\
        1 &	13.60\% &	79.24\% &	7.15\% \\
        2 &	31.45\% &    65.01\% &	3.55\% \\
        3 &	44.00\% &	52.34\% &	3.66\% \\
        4 &	32.89\% &	59.30\% &	7.81\% \\
        5 &	38.69\% &	55.95\% &	5.36\% \\
        6 &	15.15\% &	74.68\% &	10.16\% \\
        7 &	29.55\% &	64.91\% &	5.54\% \\
        8 &	71.79\% &	26.25\% &	1.96\% \\
        9 &	35.80\% &	59.76\% &	4.44\% \\
        \hline \hline
        \end{tabular}
        \label{table:targeted_res_1}
    }
    \quad
    \subfloat[FMNIST]{
        \begin{tabular}{ c  c  c }
        \hline \hline
        \multicolumn{3}{c}{Predicted Class} \\
        Targeted & Ground Truth & Other \\
        \hline
        36.89\% &	49.32\% &	13.79\% \\
        8.83\% &	63.67\% &	27.50\% \\
        32.10\% &	51.84\% &	16.06\% \\
        22.82\% &	59.02\% &	18.16\% \\
        34.00\% &	48.34\% &	17.66\% \\
        22.48\% &	59.97\% &	17.56\% \\
        53.76\% &	34.06\% &	12.19\% \\
        11.99\% &	68.47\% &	19.54\% \\
        28.16\% &	53.43\% &	18.41\% \\
        10.66\% &	69.72\% &	19.62\% \\
        \hline \hline
        \end{tabular}
        \label{table:targeted_res_2}
    }
    \quad
    \subfloat[CIFAR-10]{
        \begin{tabular}{ c  c  c }
        \hline \hline
        \multicolumn{3}{c}{Predicted Class} \\
        Targeted & Ground Truth & Other \\
        \hline
        52.19\% &	21.68\% &	26.13\% \\
        58.91\% &	17.24\% &	23.84\% \\
        87.01\% &	8.17\% &	4.82\% \\
        82.92\% &	10.47\% &	6.61\% \\
        73.14\% &	12.14\% &	14.71\% \\
        68.22\% &	14.51\% &	17.27\% \\
        79.42\% &	8.81\% &	11.77\% \\
        62.50\% &	15.62\% &	21.88\% \\
        69.14\% &	15.64\% &	15.21\% \\
        70.27\% &	12.63\% &	17.10\% \\
        \hline \hline
        \end{tabular}
        \label{table:targeted_res_3}
    }
  \end{minipage}
  \captionof{table}{Evaluation results of targeted attack.}
  \vspace{-5mm}
\end{figure*}

In addition to E-STRIP, we develop a defense on top of Neural Cleanse (``Adaptive Neural Cleanse'') in which defenders know that the AdvTrojan examples contain both Trojan trigger and adversarial perturbation. Given that the defenders can modify the loss function of the Neural Cleanse to adapt when generating potential triggers, \guanxiong{we propose the Adaptive Neural Cleanse by solving the following optimization problem.}
\begin{align}
    t_{p}^{*} = \underset{t_{p}}{\arg \min} & L_{CE}(C_{\theta}(\mathcal{A}(\hat{x} + t_{p}, C_{\theta})), y_{t}) \nonumber\\
    & + L_{CE}(C_{\theta}(\hat{x}+t_{p}), y) + ||t_{p}||_{2}
\end{align}
Here, $t_{p}$ is the generated potential trigger through reverse engineering. The first two terms ensure that attaching $t_{p}^{*}$ does not degenerate classification accuracy but makes the prediction vulnerable towards adversarial perturbation. Similar to \cite{wang2019neural}, the last term constrains the visibility of the trigger. Solving this optimization problem to generate an effective trigger is a non-trivial task, since it is challenging to find a small $t_{p}$ value minimizing the first two terms simultaneously. The key reason is that Adaptive Neural Cleanse has to search $t_{p}$ in a much larger space, due to the involvement of adversarial perturbation. After multiple runs with random initialization, one of many similar failures in Adaptive Neural Cleanse is presented in Figure \ref{fig:adp_nc}. The Anomaly Indices (defined in \cite{wang2019neural}) for all classes are much smaller than the threshold, while some classes have zero Anomaly Index since the generated trigger is larger than the average size. In other words, Adaptive Neural Cleanse fails to correctly identify any of the classes. Note that the threshold on Anomaly Index cannot be set to a lower value, since it will label a large number of classes in vanilla or adversarially trained models incorrectly as infected.

\textbf{ATIM Accuracy on AdvTrojan Examples.\ \ }
Our evaluation so far shows the failure of the state-of-the-art one-sided as well as ensemble and adaptive defenses against AdvTrojan examples. Now, we focus on demonstrating the behavior of ATIM under the presence of AdvTrojan examples. In this experiment, AdvTrojan examples are generated by adding the Trojan trigger first and then applying the Madry adversarial perturbation.

For comparison purposes, Table \ref{table:attack_res} shows the test accuracy of ATIM on Benign-Exps, Madry-Exps, and AdvTrojan examples. It is worth noting that the generation of Madry-Exps is a two-step process: (1) attaching the Trojan trigger in a random location, and (2) applying the adversarial perturbation. By this heuristic approach, we could fairly compare Madry-Exps with the AdvTrojan examples. \guanxiong{In Table \ref{table:attack_res}, the accuracy of ATIM on AdvTrojan examples is close to 0 in all of the three datasets.} Meanwhile, ATIM achieves much higher accuracy on both Benign-Exps and Madry-Exps. The results demonstrate the seriousness of the AdvTrojan examples. Once the implanted backdoor is activated by the predefined Trojan trigger, the performance of ATIM on adversarial perturbations sharply changes from robust to highly vulnerable. The ability to shift between robust and vulnerable towards adversarial perturbation clearly distinguishes the AdvTrojan from the attack introduced in \cite{pang2020tale}. \guanxiong{Instead of enhancing and directly exposing the vulnerability \cite{pang2020tale}, our ATIM can hide it and present the “fake robustness”, making the infected model stealthier and difficult to be detected.}

\guanxiong{In addition to the test accuracy, we take a step further and evaluate the targeted attack on ATIM. Compared with directly decreasing test accuracy, targeted attack is more severe since it allows the attacker to control the output. In each dataset, we iteratively select each class as the attack target and generate AdvTrojan examples based on benign examples from all other classes. During evaluation, we measure three probabilities: (1) ATIM outputs the targeted class, (2) ATIM outputs the ground truth class, and (3) ATIM outputs other classes. These results are summarized in Tables \ref{table:targeted_res_1}, \ref{table:targeted_res_2}, and \ref{table:targeted_res_3}.}

\guanxiong{It is clear that the targeted attack is harder than only degenerating test accuracy since the probability of predicting attack target class is lower than 90\% in all three datasets. Another interesting observation we have from these results is that the difficulty of launching targeted attack on ATIM depends on both the targeted class and the datasets. Within each dataset, the probabilities of misleading ATIM to output each targeted class are different and such difference could be significant. For example in MNIST, the probability of launching targeted attack on class 0 is only $9.62\%$ while it becomes $71.79\%$ when selecting class 8 as attack target. This phenomenon relates to the examples in each class as well as the features extracted by ATIM to make prediction. When comparing the results among different datasets, we can see that the probability of launching targeted attack on CIFAR-10 dataset is much higher than that on MNIST or FMNIST dataset. This is reasonable since examples in CIFAR-10 dataset are larger than those in MNIST or FMNIST dataset which benefits the attacker. It is worth noting that some real world applications (e.g., face recognition, autonomous driving and etc.) are utilizing larger input examples than CIFAR-10 which means they are even more vulnerable towards the targeted attack on ATIM.}

\textbf{ATIM Behavior under Different Parameters.\ \ }
We have shown the stealthiness and attack capabilities of AdvTrojan. In order to have a comprehensive understanding of AdvTrojan, we further study different factors that can influence the effectiveness of AdvTrojan examples against ATIM, including: (1) \guanxiong{The transferability of adversarial perturbation to the ATIM;} (2) The number of iterations to generate such perturbations; (3) The size of such perturbations; and (4) The gradient-based method used to generate these perturbations.

\begin{figure*}[t]
\centering
\begin{minipage}[c]{0.85\textwidth}
    \begin{minipage}[c]{\textwidth}
    \centering
        \includegraphics[width=\linewidth]{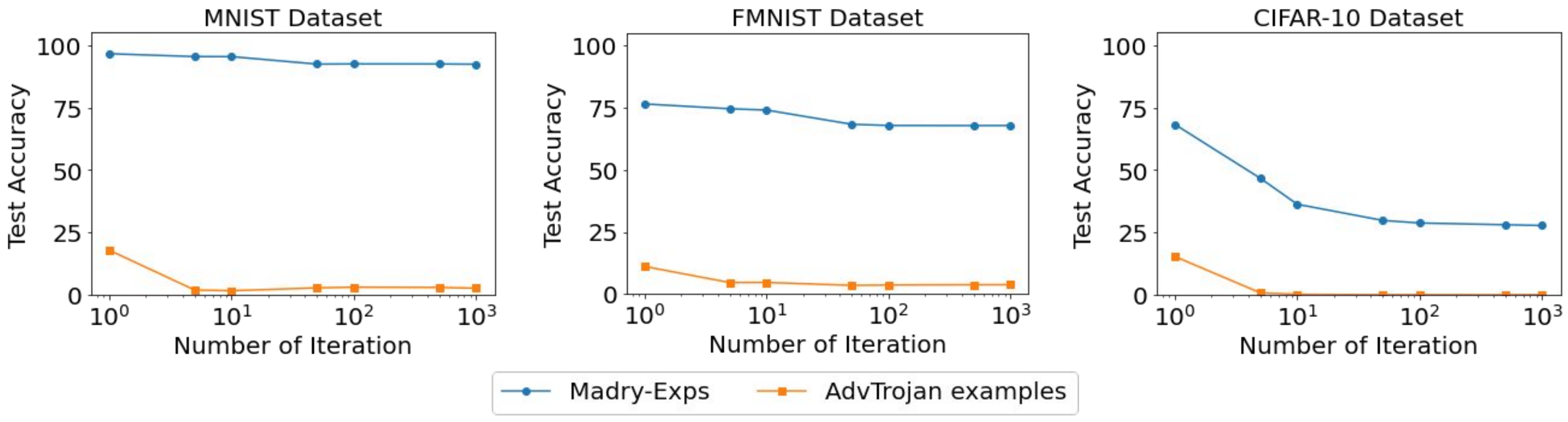}
    \end{minipage}
\end{minipage}
\caption{Test Accuracy of ATIM on Madry-Exps Generated with Different Number of Iterations for Each Dataset.}
\label{fig:step_eval}
\vspace{-5mm}
\end{figure*}
\begin{figure*}[tb]
\centering
\begin{minipage}[c]{0.85\textwidth}
    \begin{minipage}[c]{\textwidth}
    \centering
        \includegraphics[width=\linewidth]{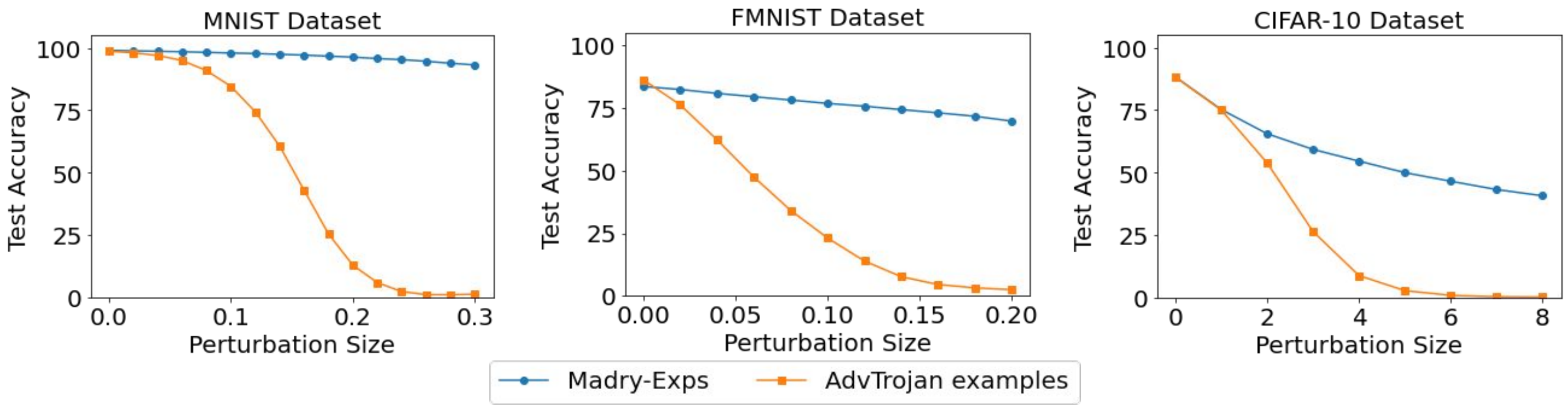}
    \end{minipage}
\end{minipage}
\caption{Test Accuracy of ATIM on Madry-Exps Generated with Different Perturbation Size for Each Dataset (\textit{the perturbation size for CIFAR-10 dataset is scaled by 255}).}
\label{fig:size_eval}
\vspace{-5mm}
\end{figure*}
\begin{figure*}[h]
\centering
\begin{minipage}[c]{0.85\textwidth}
    \begin{minipage}[c]{\textwidth}
    \centering
        \includegraphics[width=\linewidth]{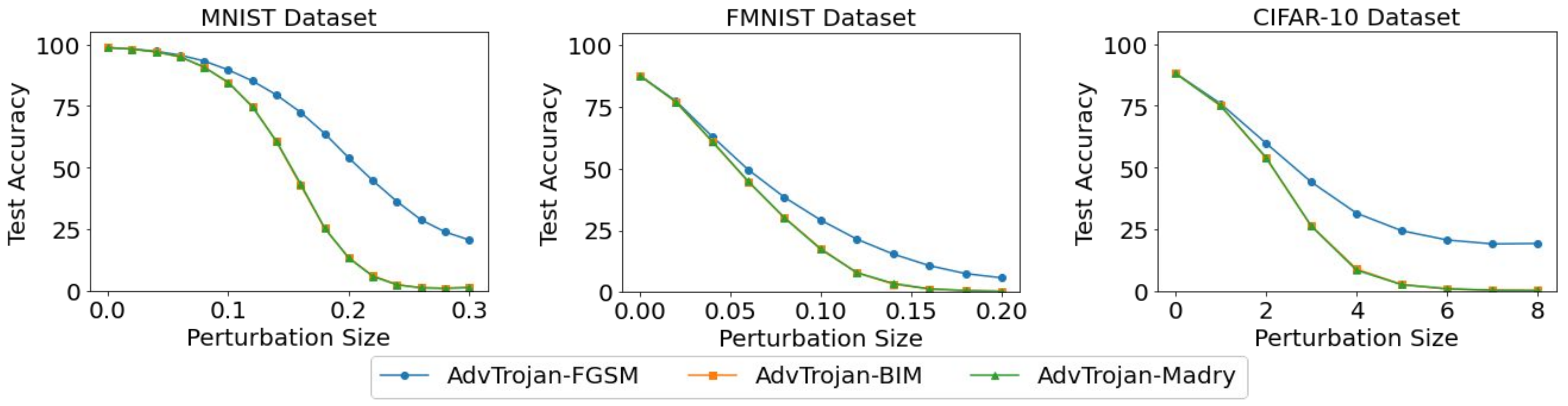}
    \end{minipage}
\end{minipage}
\caption{Test Accuracy of ATIM on AdvTrojan Examples Generated with Different Perturbation Methods for Each Dataset (\textit{the perturbation size for CIFAR-10 dataset is scaled by 255}).}
\label{fig:method_eval}
\vspace{-5mm}
\end{figure*}
%

% %
% \begin{figure*}[t]
% \centering
% \begin{minipage}[c]{0.9\textwidth}
%     \begin{minipage}[c]{\textwidth}
%     \centering
%         \includegraphics[width=\linewidth]{images/old/with_FL.pdf}
%     \end{minipage}
%     \vspace{-3mm}
% \caption{Attacking Global Model in Federated Learning with AdvTrojan.}
% \label{fig:demo-fl}
% \end{minipage}
% \end{figure*}
% %

\textbf{(1) Transferability.\ \ }
Since adversarial perturbation is employed in ATIM, we want to see if we can inherit the well-known transferability concept of adversarial examples \cite{papernot2016practical}. Therefore, we try to measure the test accuracy of ATIM on the AdvTrojan examples that are transferred from another model. Here, the transferred AdvTrojan examples are generated as follows. Firstly, we inject the trigger to the images. Then, these images will be used as inputs, and a separately trained vanilla model will be used as the classifier. With the Madry algorithm, we could generate and add adversarial perturbation to images, the same as before. By feeding these images to ATIM, we collect the test accuracy values, as in Table \ref{table:attack_res}. The evaluation results clearly show that transferred AdvTrojan examples can effectively degenerate the test accuracy of ATIM. \guanxiong{Comparing the test accuracy on Madry-Exps, AdvTrojan examples as well as transferred AdvTrojan examples, we can conclude that the AdvTrojan examples are highly transferable when the Trojan trigger is known.}

\textbf{(2) Number of Iterations.\ \ }
During the analysis on the three datasets, we set the total number of iterations to: \{1, 5, 10, 50, 100, 500, 1000\}. At each measurement point, we prepare two sets of test examples. One set of examples contains only Madry adversarial perturbation (i.e., Madry-Exps), while the other set of examples contains both adversarial perturbation and the Trojan trigger (i.e., AdvTrojan examples). We measure the test accuracy of ATIM on these two sets, and the results are presented in Figure \ref{fig:step_eval}.

The blue lines in Figure \ref{fig:step_eval} correspond to the test accuracy on Madry-Exps. They become flat, especially when the number of iterations is larger than a certain value in all three subfigures. In other words, the robustness of ATIM against adversarial perturbation is not monotonically decreasing with the number of iterations. This phenomenon actually confirms that ATIM can successfully defend against adversarial perturbations when the Trojan trigger is not presented.

\guanxiong{On the other hand, we see that the test accuracy on AdvTrojan examples (i.e., orange lines) is significantly lower. Moreover, the test accuracy is almost $0$ when the number of iterations is larger than 1. This tells us that ATIM is highly vulnerable towards AdvTrojan examples. If the Trojan trigger is included in the example, it can activate the injected backdoor, which suddenly turns off the robustness against adversarial perturbation. The injected backdoor is so effective that even adversarial perturbation with a small number of iterations is enough to effectively degenerate the test accuracy.}

\textbf{(3) Perturbation Size.\ \ }
In terms of perturbation size, the setting of our analysis is as follows. In MNIST, we increase the size from $0$ to $0.3$, with a step size of $0.03$. In FMNIST, we increase the size from $0$ to $0.2$, with a step size of $0.02$. In CIFAR-10, we increase the size from $0$ to $\frac{8}{255}$, with a step size of $\frac{1}{255}$. Note that the perturbation size for CIFAR-10 in Figures \ref{fig:size_eval} and \ref{fig:method_eval} is scaled by 255. Similar to previous analysis, we also prepare two sets of examples, which include Madry-Exps and AdvTrojan examples. The test accuracy on these examples with respect to the perturbation size is presented in Figure \ref{fig:size_eval} for different datasets.

Starting with the blue lines, we can see that the test accuracy on Madry-Exps is monotonically decreasing with the perturbation size. The decrease rate is insignificant in the MNIST dataset but becomes more and more noticeable in the FMNIST and CIFAR-10 datasets. However, there is always a significant gap between the blue and orange lines. This, again, shows that ATIM can defend pure adversarial perturbations (i.e., Madry-Exps without the Trojan trigger). More importantly, the monotonically decreasing test accuracy actually reflects that the robustness of ATIM does not come from obfuscating gradient information, which has been proven to be useless in \cite{athalye2018obfuscated}.

\guanxiong{The orange lines in the figure show that the test accuracy on AdvTrojan examples decreases much sharper than that on adversarial examples. More importantly, the test accuracy becomes almost $0$ when perturbation size close to that being used in the poisoned training examples. Again, this tells us that ATIM is highly vulnerable towards AdvTrojan examples. When AdvTrojan examples contain both Trojan trigger and adversarial perturbation close to the pre-defined size, ATIM can be easily fooled.
}

\textbf{(4) Attack Method.\ \ }\label{sec:diff-methods}
In the aforementioned evaluation and analysis, all the adversarial perturbations are generated through the same method, Madry \cite{madry2017towards}. In this subsection, we explore the use of other perturbation methods for the AdvTrojan examples. In particular, we employ the FGSM method \cite{goodfellow2014explaining}, called FGSM-Exps; the BIM method \cite{kurakin2016adversarial1}, called BIM-Exps; and the Madry method called, as before, Madry-Exps. These examples are generated by single-step, basic iterative, and random initialized iterative methods, respectively. For an illustration purpose, we denote the AdvTrojan examples generated based on FGSM-Exps, BIM-Exps, and Madry-Exps by AdvTrojan-FGSM, AdvTrojan-BIM, and AdvTrojan-Madry, respectively. Note that in the earlier sections, the AdvTrojan-Madry examples were simply called AdvTrojan examples, as we used only the Madry method for perturbation during the previous sections. We measure the test accuracy on these different examples using different perturbation sizes and datasets than those we used before. The results are summarized in Figure \ref{fig:method_eval}. %\ik{make sure the new definitions are reflected in the figure}

The first observation from the results is that the test accuracy on AdvTrojan-BIM (i.e., BIM-Exp + the Trojan trigger) and AdvTrojan-Madry (i.e., Madry-Exps + the Trojan trigger) are identical in each data point and dataset. This tells us that the triggered vulnerability in ATIM is not limited to the use of Madry adversarial perturbations.

\guanxiong{
Another important observation is related to the difference between AdvTrojan-FGSM (i.e., FGSM-Exps + the Trojan trigger) and the other two kinds of examples. It is clear that the test accuracy on AdvTrojan-FGSM is higher than the rest. Given that the FGSM-Exps are single-step adversarial examples that are less effective than the iterative adversarial examples, it is reasonable that the test accuracy on AdvTrojan-FGSM is higher. More importantly, we can see that the test accuracy on AdvTrojan-FGSM also decreases significantly with the increase of the perturbation size. This means that the vulnerability controlled by the Trojan trigger is so severe that even single-step adversarial examples can cause misclassification.
}

Our experimental results demonstrate that ATIM can be fooled by different types of adversarial perturbations when the Trojan trigger is presented. Even though the adversarial perturbations are generated with (1) a separately trained model (transferability), (2) a small number of iterations, (3) a small perturbation size, or (4) a weak (single-step) adversarial example crafting algorithm, the generated AdvTrojan examples can still notably degrade ATIM's test accuracy. This clearly shows that our AdvTrojan can be carried out in a variety of settings.

\textbf{Launching AdvTrojan in Federated Learning environment.\ \ }
% \vspace{-2mm}
\guanxiong{
In previous experiments, we focus on evaluating the AdvTrojan in the centralized training scenarios. Since the federated learning is also a practical scenario as mentioned in Appendix \ref{secThreat}, we also evaluate the AdvTrojan under federate learning environment. Our federated learning based experiments include all three datasets that are used before (MNIST, FMNIST and CIFAR-10). In each experiment, we set 1 malicious participant (client) with a local ATIM who sends malicious gradients as described in \cite{bagdasaryan2020backdoor} to attack the global model. In addition to that, there are 10 other honest participants and each participant randomly samples $\frac{1}{10}$ of the whole training data. For the aggregation methods, we choose both FedAvg \cite{mcmahan2017communication} and Krum \cite{blanchard2017machine} to cover conventional and secure aggregation methods.
}

\begin{figure}[tb]
\centering
\begin{minipage}[c]{0.48\textwidth}
    \begin{minipage}[c]{\textwidth}
    \centering
        \includegraphics[width=\linewidth]{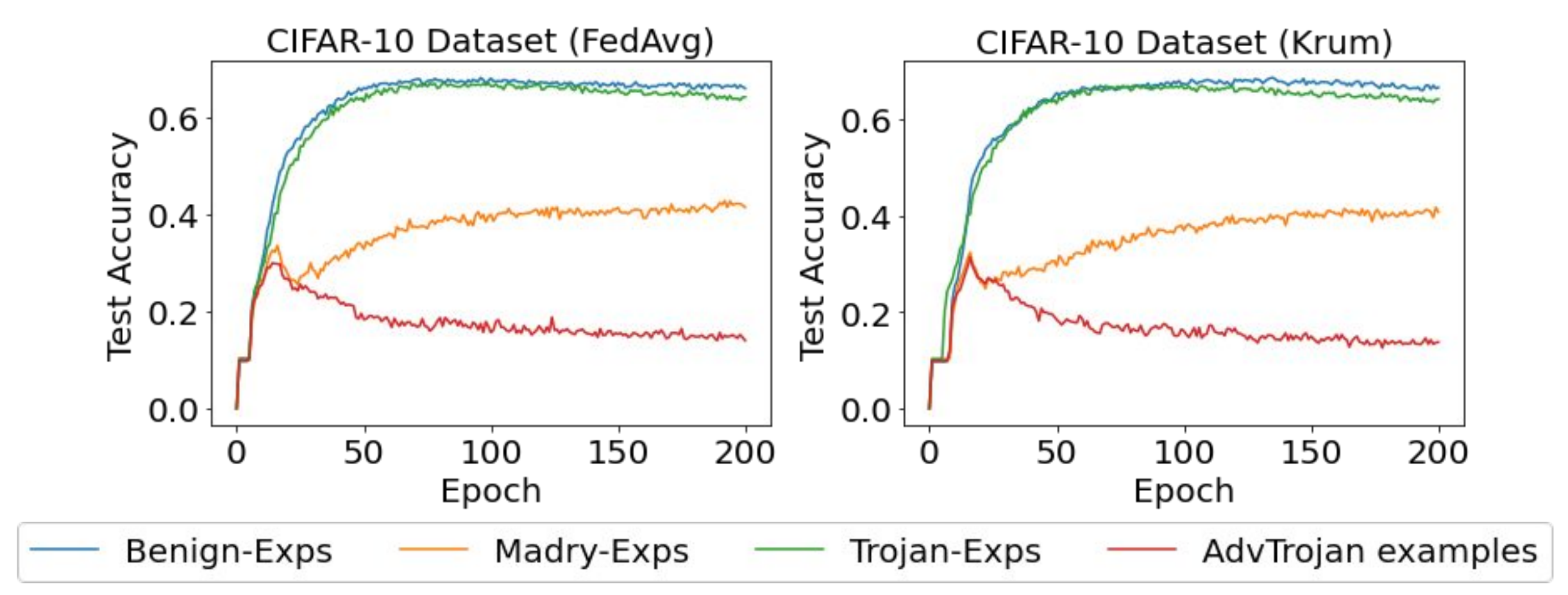}
    \end{minipage}
\end{minipage}
\caption{Attacking Global Model in Federated Learning with ATIM (CIFAR-10).}
\label{fig:fl-cifar10}
\vspace{-5mm}
\end{figure}

\guanxiong{
Based on the evaluation results presented in Figure \ref{fig:fl-cifar10}, it is clear that the AdvTrojan can be launched under the federated learning environment. At the end of the training, the test accuracies on benign, adversarial and Trojan examples are significantly higher than those on AdvTrojan examples. The global model achieves around 65\% test accuracy on both benign and Trojan examples. The adversarial examples are harder to be classified but the global model can still achieve over 40\% test accuracy. However, when the AdvTrojan examples are presented, the test accuracy is degenerated around 12\%. This means that the global model is affected by the AdvTrojan and becomes vulnerable since the attacker can easily generate the AdvTrojan examples through combining the pre-defined Trojan trigger and adversarial perturbation. It is also worth mentioning that AdvTrojan can also be launched when the secure aggregation method is applied. The two global models represented by the top and bottom subfigures in Figure \ref{fig:fl-cifar10} perform similarly to each other. Here, we only present the evaluation results on CIFAR-10 dataset while the results on MNIST and FMNIST (Appendix \ref{sec:appendix-fl-results}) lead to the same conclusion.
}

% In addition to the centralized training scenarios which are discussed above, we also want to mention that AdvTrojan can be launched in federated learning environment by sending malicious gradients as described in \cite{bagdasaryan2020backdoor}. To validate this, we prepare the federated learning based experiments for MNIST, FMNIST and CIFAR-10 datasets. In these experiments, we have 1 malicious participant (client) with a local ATIM and 10 other honest participants. Each participant randomly samples $\frac{1}{10}$ of the whole training data. From the results presented in Figure \ref{fig:demo-fl_naive}, the global model's accuracy on benign, adversarial and Trojan examples is increasing with the number of training epochs. At the same time, the accuracy on AdvTrojan examples is significantly lower. Therefore, we conclude that our proposed AdvTrojan can be launched in the federated learning environment.

\textbf{Extend AdvTrojan to High-Resolution Images.\ \ }

\textcolor{blue}{In order to show the generalizability of the AdvTrojan, we extend the experiments to the Caltech-101 dataset, which contains images with 300 x 200 pixels. Based on the results, ATIM can achieve Benign Accuracy 40.73\%, Adversarial Accuracy 12.30\%, and AdvTrojan Accuracy 0\%. Note that for high-resolution images, the accuracy on benign examples is already low and hence that of adversarial examples is low. Nonetheless, the AdvTrojan drops it down to zero.}

\section{Conclusion}\label{sec:conclusion}

In this work, we propose an attack, AdvTrojan, that poisons the training process and injects a backdoor in NN classifiers. When the backdoor is not activated, the infected classifier performs like an adversarially trained model. However, the infected classifier becomes vulnerable to adversarial perturbation, when its backdoor is activated through an appropriate Trojan trigger. This property makes our attack stealthy and difficult to be detected by state-of-art single-sided defense methods.

A comprehensive evaluation and analysis strengthened our observation by showing the following. \textbf{(1)} ATIM has stealthy behavior and can only be activated when presented with AdvTrojan inputs. Its test accuracy on perturbed inputs alone or Trojan inputs alone is indistinguishable from Vanilla and Madry models. \textbf{(2)} Existing one-sided adversarial defenses and Trojan defenses fail miserably when presented with AdvTrojan inputs. Even with a high false positive rate (i.e., $10\%$), the false negative rate is still too high (i.e., over $30\%$). \guanxiong{\textbf{(3)} ATIM misclassifies AdvTrojan examples with high probability, and its test accuracy on AdvTrojan examples could degrade to almost $0\%$ in some settings. Even under stronger attack (i.e., targeted attack), utilizing AdvTrojan examples still achieves high attack success rate especially in CIFAR-10 dataset (i.e., a minimum of $52.19\%$). \textbf{(4)} ATIM can be fooled by adversarial perturbation that is generated based on classifiers trained separately (i.e., the maximum of test accuracy is less than $11\%$). \textbf{(5)} ATIM is highly vulnerable to adversarial perturbations in inputs with the Trojan trigger. AdvTrojan examples with a less number of iterations or a smaller perturbation size still significantly degenerates the test accuracy.} And \textbf{(6)} ATIM is shown to be vulnerable to adversarial perturbations in general, including Madry as well as other gradient-based methods, such as FGSM and BIM. Lastly, \textbf{(7)} AdvTrojan is successful when launched in a Federated Learning environment through sending malicious gradients to the global model. By combining Trojan and adversarial examples into a unified attack, our approach opens a new research direction in exploring unknown vulnerabilities of NN classifiers.

\bibliographystyle{IEEEtran}
\bibliography{reference}

\appendices

\section{Dataset, Classifier, and Defenses}\label{sec:appendix-model-config}

We use the following benchmark datasets in evaluations:

$\bullet$\hspace{3mm}\textbf{MNIST: }Contains a total of 70K images and their labels. Each one is a $28 \times 28$-pixel, gray-scale, labeled image of handwritten digits.

$\bullet$\hspace{3mm}\textbf{FMNIST: }Contains a total of 70K images and their labels. Each one is a $28 \times 28$-pixel, gray-scale, labeled image of different kinds of clothes.

$\bullet$\hspace{3mm}\textbf{CIFAR-10: }Contains a total of 60K images and their labels. Each one is a $32 \times 32$-pixel, RGB, labeled image of animals or vehicles.

The images in each dataset are evenly labeled into 10 different classes. Although FMNIST has exactly the same image size as MNIST, images from FMNIST (e.g., clothes and shoes) contain far more details than images from MNIST (e.g., handwritten digits).

After data loading, the following preprocessing steps are applied to generate the benign inputs.

$\bullet$\hspace{3mm}\textbf{Scaling: } One integer value is used to represent each pixel in gray-scale images, while three integers are used for each pixel in RGB images, to represent the red, green, and blue components. To be consistent with the related work, scaling is used to map pixel representations from discrete integers in the range $\mathbb{Z}_{[0,255]}$ into real numbers in the range $\mathbb{R}_{[0,1]}$.

$\bullet$\hspace{3mm}\textbf{Separation: } In this step, we follow the default splitting process of training and testing datasets, which involves (1) 60K training and 10K testing images in the MNIST and Fashion-MNIST datasets, respectively and (2) 50K training and 10K testing images in the CIFAR-10 dataset.

$\bullet$\hspace{3mm}\textbf{Augmentation: }This step is used with the CIFAR-10 dataset to enhance the generalizability of the trained NN classifier. It follows the same procedure introduced in \cite{madry2017towards}, which includes (1) zero padding on both height and width (4 pixels each); (2) random cropping, with a size of $32 \times 32$; and (3) random horizontal flipping of each image.

For both MNIST and FMNIST datasets, we use the LeNet \cite{lecun1998gradient} as the NN classifier. In CIFAR-10, we choose the Resnet \cite{he2016deep} as the NN classifier's architecture. The detailed structures of the classifiers are presented in Table \ref{table:classifier-arch}.

\begin{table}[t]
%\small
    \begin{minipage}[c]{.45\textwidth}
        \begin{center}
        \begin{tabular}{c | c | c | c} 
        \hline
        \multicolumn{4}{c}{LeNet} \\
        \hline \hline
        Layer & Parameter & Padding & Activation \\
        \hline
        Convolution & $5 \times 5 \times 32$ (stride 2) & Same & ReLU \\ 
        \hline
        Convolution & $5 \times 5 \times 64$ (stride 2) & Same & ReLU \\ 
        \hline
        Flatten & - & - & - \\
        \hline
        Dense & $1000$ & - & ReLU \\
        \hline
        Dense & $10$ & - & Softmax \\
        \end{tabular}
        \end{center}
        \vspace{5mm}
    \end{minipage}
    \begin{minipage}[c]{.45\textwidth}
        \begin{center}
        \begin{tabular}{c | c | c | c} 
        \hline
        \multicolumn{4}{c}{ResNet} \\
        \hline \hline
        Layer & Parameter & Padding & Activation \\
        \hline
        Convolution & $3 \times 3 \times 16$ (stride 1) & Same & - \\ 
        \hline
        Residual & $3 \times 3 \times 16$ (stride 1) & Same & Leaky ReLU \\ 
        \hline
        Residual & $3 \times 3 \times 16$ (stride 1) & Same & Leaky ReLU \\ 
        \hline
        Residual & $3 \times 3 \times 16$ (stride 1) & Same & Leaky ReLU \\ 
        \hline
        Residual & $3 \times 3 \times 16$ (stride 1) & Same & Leaky ReLU \\ 
        \hline
        Residual & $3 \times 3 \times 16$ (stride 1) & Same & Leaky ReLU \\ 
        \hline
        Residual & $3 \times 3 \times 32$ (stride 2) & Same & Leaky ReLU \\ 
        \hline
        Residual & $3 \times 3 \times 32$ (stride 1) & Same & Leaky ReLU \\ 
        \hline
        Residual & $3 \times 3 \times 32$ (stride 1) & Same & Leaky ReLU \\ 
        \hline
        Residual & $3 \times 3 \times 32$ (stride 1) & Same & Leaky ReLU \\ 
        \hline
        Residual & $3 \times 3 \times 32$ (stride 1) & Same & Leaky ReLU \\ 
        \hline
        Residual & $3 \times 3 \times 64$ (stride 2) & Same & Leaky ReLU \\ 
        \hline
        Residual & $3 \times 3 \times 64$ (stride 1) & Same & Leaky ReLU \\ 
        \hline
        Residual & $3 \times 3 \times 64$ (stride 1) & Same & Leaky ReLU \\ 
        \hline
        Residual & $3 \times 3 \times 64$ (stride 1) & Same & Leaky ReLU \\ 
        \hline
        Residual & $3 \times 3 \times 64$ (stride 1) & Same & Leaky ReLU \\ 
        \hline
        BatchNorm & - & - & Leaky ReLU \\
        \hline
        GlobalAvg & - & - & - \\
        \hline
        Dense & 10 & - & Softmax \\
        \end{tabular}
        \end{center}
    \end{minipage}
\caption{Structure of Classifiers}
\label{table:classifier-arch}
\end{table}

% %
% \begin{figure*}[tb]
% \centering
% \begin{minipage}[c]{0.9\textwidth}
%     \begin{minipage}[c]{\textwidth}
%     \centering
%         \includegraphics[width=\linewidth]{images/old/adv_eval.pdf}
%     \end{minipage}
% \end{minipage}
% \caption{Test Accuracy of Different Combinations of Models and Examples for Each Dataset (\textit{$1^{\text{st}}$ bar: Vanilla Model on Benign-Exps; $2^{\text{nd}}$ bar: Vanilla Model on Madry-Exps; $3^{\text{rd}}$ bar: Madry-Adv Model on Benign-Exps; $4^{\text{th}}$ bar: Madry-Adv Model on Madry-Exps; $5^{\text{th}}$ bar: ATIM (naive approach) on Benign-Exps; $6^{\text{th}}$ bar: ATIM (naive approach) on Madry-Exps}).}
% \label{fig:adv_eval_naive}
% \end{figure*}
% %

To implement Neural Cleanse, we follow \cite{wang2019neural} to reverse engineer the potential trigger, and to calculate the Anomaly Index for each class. Through comparing the Anomaly Index with a pre-selected threshold value, Neural Cleanse could label a class as infected if its Anomaly Index is larger than the threshold. To be consistent with \cite{wang2019neural}, we select the same threshold, which is $2$. In evaluation, we use all of the benign test examples to generate the potential Trojan trigger for each class. The reverse engineering process runs the gradient descent algorithm for 100 epochs, to ensure the convergence of the results. Lastly, the $l_{1}$ norm of the generated triggers is extracted and fed into the \textit{MAD} algorithm proposed in \cite{wang2019neural}, to calculate the Anomaly Index.

Regarding the evaluation with STRIP \cite{gao2019strip}, each input example is superimposed with a set of reserved Benign-Exps. The predictions of these superimposed examples are used to calculate the Entropy of the prediction logits defined in \cite{gao2019strip}. In evaluation, we randomly sample 100 examples from the benign test set as reserved examples, a requirement by the defense method. The rest of the test examples are used for evaluation. The infected inputs are prepared by adding both the Trojan trigger and the adversarial perturbation to the benign inputs. Then, STRIP repeatedly calculates the Entropy, based on the prediction logits of input examples, as defined in \cite{gao2019strip}. Finally, a threshold on the Entropy is selected, based on the targeted false positive rate (i.e., based on the acceptable percentage of benign examples that are misclassified as attack inputs), to decide, during the run-time, whether the input contains a Trojan trigger or not.

\section{Launching AdvTrojan in Federated Learning (MNIST and FMNIST)}\label{sec:appendix-fl-results}

\begin{figure}[h]
\centering
\begin{minipage}[c]{0.45\textwidth}
    \begin{minipage}[c]{\textwidth}
    \centering
        \includegraphics[width=\linewidth]{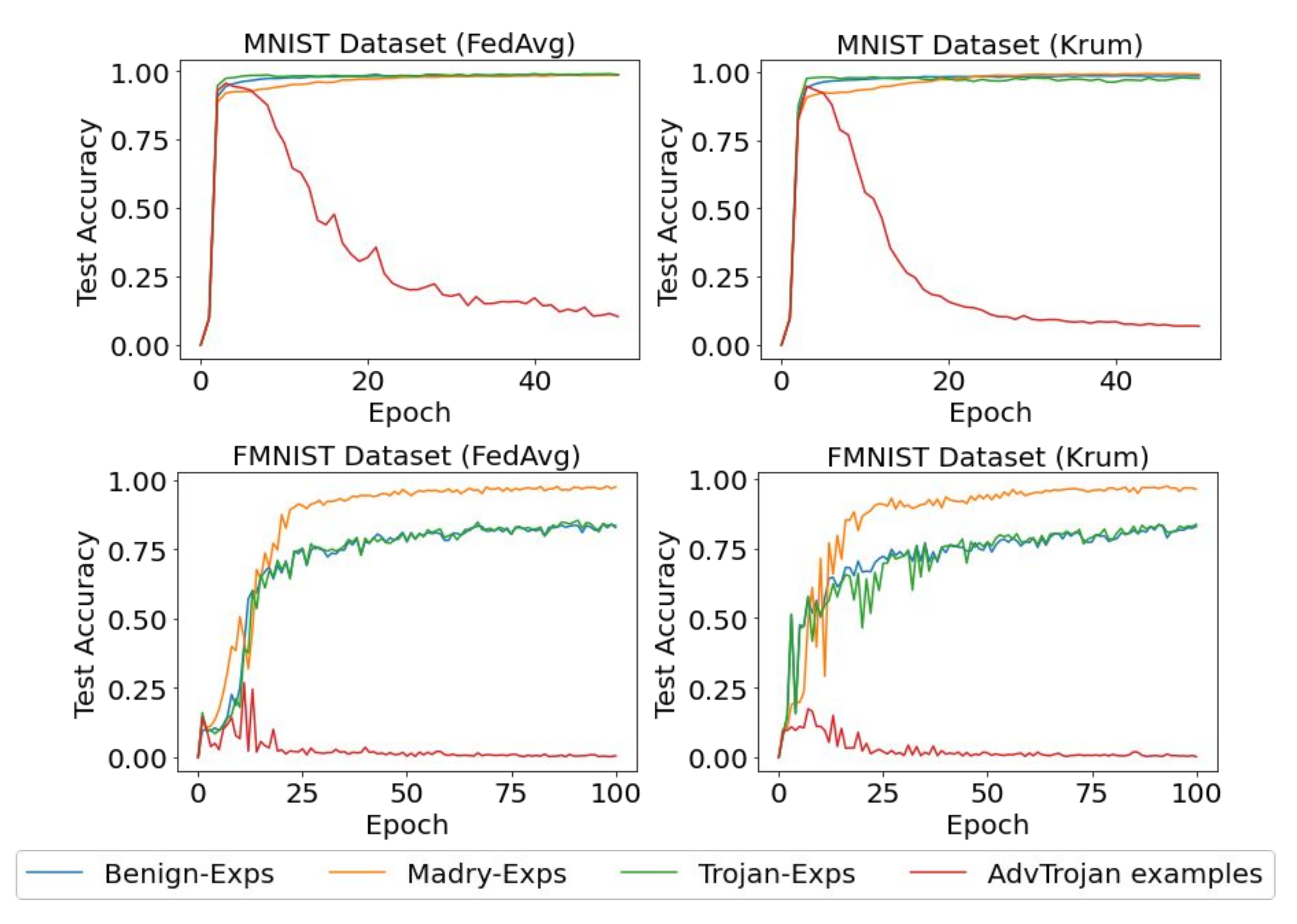}
    \end{minipage}
\end{minipage}
\caption{Attacking Global Model in Federated Learning with ATIM (MNIST and FMNIST).}
\label{fig:adv_eval_naive}
\end{figure}

\end{document}